\documentclass[prl,twocolumn,showpacs,nofootinbib,preprintnumbers,amsmath,amssymb,aps,longbibliography,superscriptaddress,floatfix]{revtex4-2}
\usepackage{graphicx,subfigure,epsfig}
\usepackage{dcolumn}
\usepackage{amssymb}
\usepackage{times}
\usepackage{amsmath}
\usepackage{amsfonts}
\usepackage{mathrsfs}
\usepackage{setspace}
\usepackage{latexsym}
\usepackage{bbm}
\usepackage{float}
\usepackage{flafter}
\usepackage{bm}
\usepackage{epstopdf}
\usepackage{color}
\usepackage{multirow}
\usepackage{physics} 
\usepackage[export]{adjustbox}
\DeclareMathOperator{\sgn}{sgn}
\usepackage{braket}

\usepackage{tikz}                  

\usepackage{dsfont}

\newcommand*\diff{\mathop{}\!\mathrm{d}}

\usepackage{soul} 

\def \be {\begin{eqnarray}}
\def \ee {\end{eqnarray}}

\newcommand{\bl}{{\bf l}}

\newcommand{\bt}{{\bf t}}

\newcommand{\bphi}{{\bm{\phi}}}

\usepackage{xcolor}

\newcommand{\bea}{\begin{equation} \begin{aligned}}
\newcommand{\eea}{\end{aligned} \end{equation} }

\newcommand{\bpm}{\begin{pmatrix}}
\newcommand{\epm}{\end{pmatrix}}

\usepackage{xr}
\usepackage[pdftex,breaklinks,bookmarks=false,colorlinks,linkcolor=blue,citecolor=blue,urlcolor=blue,hypertexnames=false]{hyperref}

\usepackage[normalem]{ulem}

\usepackage{cleveref}
\crefname{appendix}{App.}{Apps.}
\crefname{equation}{Eq.}{Eqs.}
\crefname{figure}{Fig.}{Figs.}
\crefname{table}{Tab.}{Tabs.}
\crefname{section}{Sec.}{Secs.}
\creflabelformat{appendix}{#2#1#3}

\usepackage{orcidlink}

\begin{document}

\title{Quantized Transport of $\nu = 2/3$ Fractional Quantum Hall Edge \\ with Disordered Superconducting Proximity}
\author{Pok Man Tam \orcidlink{0000-0001-9620-4689}}
\thanks{These two authors contributed equally.}
\affiliation{Princeton Center for Theoretical Science, Princeton University, Princeton, New Jersey 08544, USA}
\author{Hao Chen \orcidlink{0000-0003-1084-9166}}
\thanks{These two authors contributed equally.}
\affiliation{Department of Physics, Princeton University, Princeton, New Jersey 08544, USA}
\affiliation{Department of Electrical and Computer Engineering, Princeton University, Princeton, New Jersey 08544, USA}
\author{Biao Lian}
\thanks{Email: biao@princeton.edu}
\affiliation{Department of Physics, Princeton University, Princeton, New Jersey 08544, USA}

\begin{abstract}
Quantum Hall edge states in proximity to a superconductor (SC) usually acquire a non-quantized electron-to-hole conversion probability in transport, due to non-universal SC couplings and disorders. With \textit{counter-propagating} modes, we show that the situation can be the opposite in the $\nu=2/3$ fractional quantum Hall (FQH) edge states with SC proximity, where disordered SC-couplings can reconstruct the edge states into an infinite set of stable phases with quantized electron-to-hole conversion probability along a long edge. Each phase is dominated by a disordered SC-coupling that tunnels $\pm \abs{q_N}$ Cooper pairs, which can take values $\abs{q_N}=1, 4, 15$, etc. We predict that this gives rise to a quantized downstream resistance $R_d = h/(2q^2_Ne^2)$ in an FQH-SC junction, serving as a quantized electrical transport signature beyond the Hall conductance. Higher-order nonlinear transport due to irrelevant Cooper pair tunneling or vortex dissipation is further studied, which becomes dominant when the edge is in a normal phase. Our results apply to both the single-layer state (as a particle-hole conjugate of $\nu=1/3$) and the bilayer Halperin-(112) state, revealing a rich landscape of disorder-stabilized phases in FQH edge states with SC proximity, and may as well apply to fractional Chern insulators recently observed at the same filling.

\end{abstract}
\maketitle

\noindent {\color{blue}\emph{Introduction.}}
Fractional quantum Hall (FQH) states \cite{Tsui1982} in proximity to superconductor (SC) provide a rich platform for novel phases of matter. It is predicted that the SC proximity in the bulk of FQH states may lead to non-Abelian Ising and Fibonacci anyons \cite{Lindner2012,clarke2013exotic, Mong2014_TQC, Vaezi2014,qi2010,wangj2015}, which serve as building blocks for topological quantum computing \cite{kitaev2003fault, Nayak2008_RMP, lian2018topological}. Such bulk SC proximity proposals are, however, experimentally challenged by the lack of smoking-gun signatures, and the inevitable competition between strong magnetic field and SC \cite{tinkham2004introduction}. Another important direction is to investigate quantum Hall states with SC proximity on the \emph{edge}, which is more feasible and has seen substantial developments both theoretically \cite{ma1993josephson, Fisher1994, Zyuzin1994, Hoppe2000, Giazotto2005, Akhmerov2007, Ostaay2011, Stone2011, Lian2016, Tanaka2021a, Tanaka2021b, manesco2022mechanisms, Lian2018, Lian2019, GaneshanLevin2022, kurilovich2023disorder, Schiller2023,hu2024resistance} and experimentally \cite{rickhaus2012quantum,Komatsu2012,wan2015induced, amet2016supercurrent, lee2017inducing, park2017propagation, Sahu2018, matsuo2018equal, kozuka2018andreev, seredinski2019quantum, zhao2020interference,  zhao2023loss, kim2022CAR, hatefipour2022induced, uday2024induced}. This may also be implemented in the recently observed fractional Chern insulators (FCIs) in twisted MoTe$_2$ \cite{cai2023signatures, park2023observation, zeng2023thermodynamic, PhysRevX.13.031037} and moir\'e/rhombohedral graphene \cite{Long2024fractional, Long2025extended, choi2025, xie2025,spanton2018observation, xie2021fractional}, which are topologically equivalent to FQH states \cite{PhysRevLett.106.236802, PhysRevLett.106.236803, PhysRevLett.106.236804, sheng2011fractional, PhysRevX.1.021014, Long2024fractional_review} but require no magnetic field, and the edge SC proximity may be induced either from state-of-the-art fabrication \cite{Jia2024,jia2025Sci} or from the intrinsic SC existing in the same material platform \cite{han2024signatures,litingxin2025, choi2025, ShiSenthil2025}.



\begin{figure}
    \centering
    \resizebox{0.9\columnwidth}{!}{\includegraphics[]{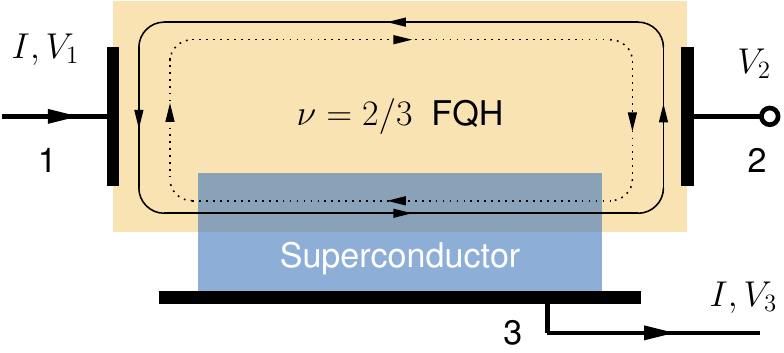}}
    \caption{The junction of $\nu=2/3$ FQH edge with SC proximity. A current $I$ is injected at lead 1 and flows out passing the SC at lead 3, while lead 2 is floating. This measures the downstream resistance $R_d\equiv V_{23}/I$ in \cref{eq: Rd,eq: Rd-non}, where $V_{ij}=V_i-V_j$, and $V_j$ is the voltage of lead $j$ ($j=1,2,3$). Note that the SC is at voltage $V_3$.
    }
    \label{fig:setup}
\end{figure}  

For quantum Hall states with SC proximity on the \emph{edge}, however, no SC-induced quantized topological transport has been predicted yet, as non-universal SC couplings and unavoidable disorders usually give non-quantized electron-to-hole conversion probabilities \cite{Lian2018, Lian2019, kurilovich2023disorder, Schiller2023,hu2024resistance,zhao2023loss}. In this Letter, we show for the first time that a robust quantized transport signature beyond the Hall conductance can be induced by disordered SC proximity on a filling $\nu=2/3$ FQH edge. The key observation is that disorders are not always destroying, but may stabilize quantized transport. A celebrated example is the $\nu=2/3$ FQH edge state without SC, as studied by Kane, Fisher and Polchinski (KFP) \cite{KFP}, where neutral impurities drive a renormalization group (RG) flow to a stable edge phase exhibiting a quantized conductance $\frac{2}{3}\frac{e^2}{h}$, in agreement with the experiments. With disordered SC proximity added to the $\nu=2/3$ FQH edge states, here we demonstrate the existence of an \textit{infinite} set of SC disorder-induced edge phases, which substantially extends the seminal analysis by KFP. Each SC edge phase is labeled by an integer $N$, and is characterized by a stable RG fixed point driven by a unique relevant disordered SC coupling that tunnels $\pm\abs{q_N} \in \mathbb{Z}$ Cooper pairs. In an FQH-SC junction illustrated in \cref{fig:setup}, we predict a \textit{quantized} downstream resistance $R_d=\frac{h}{2q_N^2 e^2}$ as an experimentally measurable fingerprint of the SC edge phase. If the proximitized edge remains in the normal phase ($q_N=0$), we show that a nonlinear current-voltage relation arises. Our theory may as well apply to the FCI found robustly at $\nu=2/3$ \cite{cai2023signatures, park2023observation, zeng2023thermodynamic, PhysRevX.13.031037, Long2024fractional, Reddy2023, Yu2024}.

{\color{blue}\emph{Clean edge.}} We begin with a review of the minimal
clean edge theory of the Abelian single-layer (or spin-polarized \cite{Yutushui2024}) $\nu=2/3$ FQH state, as a chiral Luttinger liquid with two \textit{counter-propagating} $2\pi$-compactified boson fields $\bphi=(\phi_1,\phi_2)^T$. The Lagrangian density is \cite{Wen1990composite, MacDonald1991composite, KFP, KaneFisher1995, wen1995topological, Moore1998}
\begin{align}\label{eq:2/3_Lagrangian_clean}
    \mathcal{L}_0 &= -\frac{1}{4\pi} \sum_{i,j=1}^2(\partial_t\phi_i K_{ij}\partial_x \phi_j + \partial_x\phi_i v_{ij} \partial_x\phi_j),
\end{align}
where $K_{ij}$ is the integer-valued $K$-matrix $K=\text{diag}(1,-3)$ characterizing the $\nu=2/3$ FQH topological order, and $v_{ij}$ is the positive definite velocity matrix characterizing density-density interactions. The boson fields $\bphi$ satisfy the commutation relation $[\phi_i(x), \phi_j(x')] = i\pi\text{sgn}(x-x')(K^{-1})_{ij}$ ($\text{sgn}(x)$ is the sign of $x$),  
and have a charge vector $\bt=(1,1)^T$, which gives the Hall conductance $\sigma_{H}=\bt^TK^{-1}\bt\frac{e^2}{h}=\frac{2}{3}\frac{e^2}{h}$, where $-e<0$ is the electron charge, and $h=2\pi\hbar$ is the Planck constant. 
The electric charge density is $\rho = -\frac{e}{2\pi} \bt^T \partial_x \bphi$ and the current is $j=  \frac{e}{2\pi} \bt^T \partial_t \bphi$. The vertex operator $e^{\pm i\bl^T\bphi}$ ($\bl\in\mathbb{Z}^2$ is an integer vector) annihilates/creates a particle of charge $-e\bl^T K^{-1}\bt$ and exchange statistical angle $\pi \bl^T K^{-1}\bl$.

{\color{blue}\emph{Disordered SC proximity.}}  On an edge with disordered proximity to a SC (topologically trivial and thus gives no additional edge modes) in \cref{fig:setup}, fermion-parity conserving scatterings are allowed, which are given by vertex operators $e^{i\bl^T\bphi}$ with \emph{bosonic} statistical angles $\pi \bl^T K^{-1}\bl=0\ (\text{mod }2\pi)$ \cite{Haldane1995} and \emph{even} charges $\bl^T K^{-1}\bt=2q$ ($q\in\mathbb{Z}$) (tunneling $q$ Cooper pairs). This requires $\bl=(3q+p,3q+3p)^T$, with $p\in\mathbb{Z}$ being the number of normal backscatterings. 

It is convenient to introduce a new orthonormal boson basis of an upstream ($-x$ direction) neutral mode $\phi_\sigma$ and a downstream ($+x$ direction) charged mode $\phi_\rho$:
\begin{equation}\label{eq: def charge and neutral basis}
\phi_\sigma=\sqrt{\frac{1}{2}}(\phi_1+ 3\phi_2)\ ,\quad \phi_\rho=\sqrt{\frac{3}{2}}(\phi_1+\phi_2)\ .
\end{equation}
In this basis $(\phi_\sigma,\phi_\rho)^T$, the $K$-matrix transforms into a diagonal matrix $\eta=\text{diag}(-1,1)$. 
The generic Lagrangian density with the above disordered SC proximity then becomes
\begin{equation}\label{eq:full_L_disorder}
\begin{split}
    &\mathcal{L}=\mathcal{L}_0+\mathcal{L}_\text{dis}\ ,\quad \mathcal{L}_\text{dis}=\sum_{q,p}\Big[\xi_{q,p}(x)\mathcal{O}_{q,p}+h.c.\Big]\ ,\\
    &\mathcal{O}_{q,p}\equiv e^{i(3q+p)\phi_1+i(3q+3p)\phi_2}= e^{iq\sqrt{6}\phi_\rho+ip\sqrt{2}\phi_\sigma}\ , 
\end{split}
\end{equation}
where $q\in \mathbb{Z}$, $p\in \mathbb{Z}_0^+$, and the vertex operator $\mathcal{O}_{q,p}$ annihilates charge $-2q e$ (tunnels $q$ Cooper pairs). We assume short-range correlated disorders $\xi_{q,p}(x)$ with ensemble averages $\overline{\xi_{q,p}(x)}=0$ and $\overline{\xi_{q,p}(x)\xi_{q',p'}^*(x')}=\delta_{qq'}\delta_{pp'}W_{q,p}\delta(x-x')$, and $W_{q,p}$ denotes the disorder strengths. Specifically, $\xi_{0,1}(x)$ is the non-SC disorder studied in Ref. \cite{KFP}.

{\color{blue}\emph{Stable RG fixed points.}} It can be shown \cite{GiamarchiSchulz1988, KaneFisher1995, Haldane1995,KFP, supp} that the leading order RG flow equation for the disorder strengths is $dW_{q,p}/d\lambda=(3-2\Delta_{q,p})W_{q,p}$, where $\Delta_{q,p}$ is the scaling dimension of vertex operator $\mathcal{O}_{q,p}$. Therefore, an $\mathcal{O}_{q,p}$ term is relevant (irrelevant) if $\Delta_{q,p}$ is smaller (larger) than $3/2$. (This threshold becomes $2$ for homogeneous edges with $\xi_{q,p}(x)$ being constant \cite{Haldane1995}.)

\begin{figure}[t!]
    \centering
    \includegraphics[width=\columnwidth]{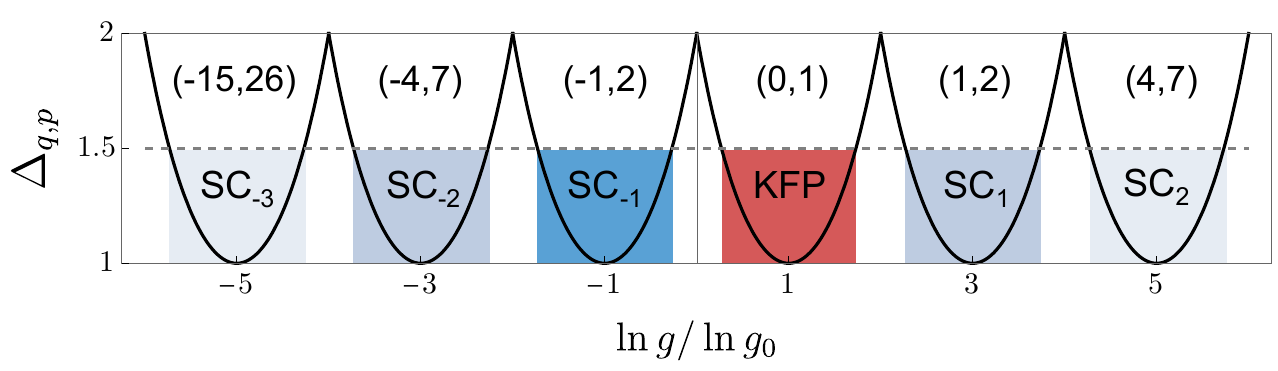}
    \caption{The black curves show the scaling dimension $\Delta_{q_N,p_N}$ of vertex operator $\mathcal{O}_{q_N,p_N}$ (label $(q_N,p_N)$ above each curve) with respect to parameter $g$ defined in Eq. \eqref{eq: def_g}, which reaches the minimum $\Delta_{q_N,p_N}^{\min}=1$ at $g_N = g_0^{2N+1}$, with $g_0=(2+\sqrt{3})^2$. Each colored interval indicates a phase SC$_N$ (with SC$_0$ being the KFP phase), in which $\mathcal{O}_{q_N,p_N}$ is relevant and $g$ flows to the RG fixed point $g_N$.}
    \label{fig:scaledim}
\end{figure}

Generically, $\mathcal{O}_{q,p}$ has a conformal dimension $h_{q,p}\ge0$ ($\bar{h}_{q,p}\ge0$) in the downstream (upstream) sector, which defines its scaling dimension $\Delta_{q,p} = h_{q,p}+\bar{h}_{q,p}$ and conformal spin $s_{q,p} = h_{q,p}- \bar{h}_{q,p}$ \cite{Haldane1995,Moore1998}. With counter-propagating modes, the scaling dimension $\Delta_{q,p}$ is non-universal and depends convexly on a single interaction parameter $g\in(0,\infty)$:
\begin{equation}\label{eq: Delta as a function of g}
    \Delta_{q,p}(g) = \frac{1}{2}\sqrt{\frac{g}{g_0}}(\sqrt{3}q-p)^2 + \frac{1}{2}\sqrt{\frac{g_0}{g}}(\sqrt{3}q+p)^2,
\end{equation}
with $g_0= (2+\sqrt{3})^2$ and $g$ a function of the positive definite matrix $v_{ij}$ (see supplementary material (SM) \cite{supp}):
\begin{equation}\label{eq: def_g}
    g=\frac{3v_{11}+v_{22}+2\sqrt{3}v_{12}}{3v_{11}+v_{22}-2\sqrt{3}v_{12}}\ .
\end{equation}
In constrast, the conformal spin $s_{q,p}=3q^2-p^2$ is a universal topological number and can be read off from \cref{eq:full_L_disorder}. This is unlike fully chiral theories where the scaling dimension and conformal spin are equal \cite{lian2019syk,huy2022}.


By definition, $\Delta_{q,p}$ is lower bounded by $\Delta^{\min}_{q,p}=|s_{q,p}|=\abs{3q^2-p^2}$, which is reachable (by \cref{eq: Delta as a function of g}). 
Since $3q^2$ cannot be a perfect square, $\Delta^{\min}_{q,p}$ cannot be $0$. This is consistent with the known fact that the $\nu=2/3$ FQH state has an ungappable edge theory (no bosonic vertex operators with zero conformal spin) \cite{Haldane1995, Moore1998, Levin2013,wangjuven2015}. Thus, an $\mathcal{O}_{q,p}$ term in \cref{eq:full_L_disorder} can be relevant only if $\Delta^{\min}_{q,p}=1$. Since $p^2, q^2 \equiv 0 \text{ or } 1 (\text{mod } 4)$, the only possibility is $p^2-3q^2=1$, which has infinite number of solutions labeled by $N\in\mathbb{Z}$ \cite{lenstra2002solving}:
\begin{equation}\label{eq:p_q_relation}
\begin{split}
    q_{N} = \frac{(2+\sqrt{3})^N -(2-\sqrt{3})^N}{2\sqrt{3}}\ ,\;\;p_N=\sqrt{3q_N^2+1}\ .
\end{split}
\end{equation}
The lowest few $q_N$ values are $0,\pm1,\pm4,\pm 15$. Figure \ref{fig:scaledim} shows $\Delta_{q_N,p_N}$ with respect to $\ln g$ (by Eq. \eqref{eq: Delta as a function of g}), which reaches its minimum $\Delta_{q_N,p_N}^{\min}=1$ when $g=g_N=(2+\sqrt{3})^{4N+2}$. 

Generically, either no vertex operator is relevant, or there is only one relevant operator $\mathcal{O}_{q_N,p_N}$ with $\Delta_{q_N,p_N}<3/2$ when the bare interaction is tuned to \cite{supp}
\begin{equation}\label{eq: condition for SC_N}
    \frac{v_{12}}{3v_{11}+v_{22}}\in\Big(B^{-}_N,B^{+}_N\Big),\;\;B^{\pm}_N = \frac{2q_{4N+2}\pm \sqrt{15}}{24 q^2_{2N+1}+18}\ .
\end{equation}
For instance, $(B^{-}_0, B^{+}_0) =(-B^{+}_{-1}, -B^{-}_{-1}) = (0.098,0.28)$. Below, we show that when $\mathcal{O}_{q_N,p_N}$ is relevant, 
its scaling dimension $\Delta_{q_N,p_N}$ flows to $\Delta^{\min}_{q_N,p_N}=1$, and the edge theory flows to $g=g_N$. This is a stable RG fixed point representing a phase of the edge theory. We denote this phase as SC$_N$, which is realized in the bare parameter range of \cref{eq: condition for SC_N}. The SC$_0$ phase is the KFP phase in the non-superconducting setting \cite{KFP}, while the prediction of SC$_{N\neq 0}$ phases is the central result of this work. 

For a relevant $\mathcal{O}_{q_N,p_N}$, we first define a \emph{fixed point boson basis} expressed in terms of $\phi_\rho$ and $\phi_\sigma$ as
\begin{equation}\label{eq: change_basis_charge_neutral}
    \begin{pmatrix}
        \phi_u \\ \phi_d
    \end{pmatrix}= \mathcal{R}_N\begin{pmatrix}
        \phi_\sigma \\ \phi_\rho
    \end{pmatrix},\quad \mathcal{R}_N=\begin{pmatrix}
      p_N & \sqrt{3}q_N \\
       \sqrt{3}q_N & p_N 
    \end{pmatrix}.
\end{equation}
For any $N\in\mathbb{Z}$, the $K$-matrix in this basis is 
$\eta=\text{diag}(-1,1)$, implying $\phi_u$ is upstream and $\phi_d$ is downstream. 
By \cref{eq:full_L_disorder}, $\mathcal{O}_{q_N,p_N}=e^{i\sqrt{2}\phi_u}$ only depends on $\phi_u$ and annihilates charge $-2q_Ne$ (for $\phi_d$, its minimal charge operator $e^{i\sqrt{\frac{2}{3}}\phi_d}$ annihilates charge $-\frac{2p_N}{3}e$).
Keeping only the relevant disorder terms, the Lagrangian density reduces to $\mathcal{L}= \mathcal{L}_d+\mathcal{L}_u+\mathcal{L}_{ud}$, with
\begin{subequations}\label{eq:full_L_general}
\begin{align}
    &\mathcal{L}_{d} = -\frac{1}{4\pi} \partial_x \phi_d(\partial_t   +v_d \partial_x )\phi_d, \\
    &\mathcal{L}_{u} = \frac{1}{4\pi} \partial_x \phi_u(\partial_t   -v_u \partial_x )\phi_u + [\xi(x) e^{i\sqrt{2}\phi_u} + h.c.],\\
    &\mathcal{L}_{ud} = -\frac{v_{ud}}{2\pi} \partial_x\phi_u \partial_x\phi_d,
\end{align}
\end{subequations}
where $\xi(x)=\xi_{q_N,p_N}(x)$, and the velocities $v_u$, $v_d$ and $v_{ud}$ are functions of $v_{ij}$ in \cref{eq:2/3_Lagrangian_clean}.

At $g=g_N$, one can show that $v_{ud}=0$, thus $\phi_u$ and $\phi_d$ decouple, suggesting this is a fixed point. Since $\Delta_{q_N,p_N}=\Delta_{q_N,p_N}^{\min}=1$ at $g=g_N$, following the method of \cite{KFP} we can redefine $e^{\pm i\sqrt{2}\phi_u}=\psi^\dag\sigma^\mp\psi$, where $\psi=(\psi_+,\psi_-)^T$ is an ancillary two-component complex fermion field of scaling dimension $1/2$, with $\psi_\pm=e^{i(\chi\pm \phi_u)/\sqrt{2}}$ defined in terms of an ancillary boson field $\chi$ identical to $\phi_u$, and $\sigma^\pm=(\sigma_x\pm i\sigma_y)/2$ are Pauli matrices. This maps $\mathcal{L}_{u}$ to a free fermion Lagrangian density $\mathcal{L}_\psi = i\psi^\dagger(\partial_t-v_u\partial_x)\psi+\psi^\dagger(\xi \sigma^-+\xi^*\sigma^+)\psi$. By a fermion basis SU(2) rotation $\widetilde{\psi}=U(x)\psi$ with $ U(x) = \mathcal{P}_x \exp\{\frac{i}{v_u}\int^{x}_{x_0} \diff{x'}\; [\xi(x')\sigma^- + \xi^*(x')\sigma^+]\}$, where $\mathcal{P}_x$ represents path ordering, one eliminates the disorder potential $\xi(x)$ and obtains $\mathcal{L}_\psi=i\widetilde{\psi}^\dagger(\partial_t-v_u\partial_x)\widetilde{\psi}$. Upon re-bosonizing $\widetilde{\psi}$ by $\widetilde{\psi}^\dag\sigma^\mp \widetilde{\psi}=e^{\pm i\sqrt{2}\widetilde{\phi}_u}$ and $\widetilde{\psi}^\dag\sigma_z\widetilde{\psi}=\frac{\partial_x\widetilde{\phi}_u}{\sqrt{2}\pi}$ 
, one arrives at the SC$_N$ fixed point Lagrangian with two free bosons $\phi_d$ (downstream) and $\widetilde{\phi}_u$ (upstream) \cite{KFP}:
\begin{equation}\label{eq:SC_fixed point_L}
    \mathcal{L}_{\text{SC}_N} = \mathcal{L}_d + \widetilde{\mathcal{L}}_u\ ,\quad \widetilde{\mathcal{L}}_u = \frac{1}{4\pi} \partial_x \widetilde{\phi}_u(\partial_t   -v_u \partial_x )\widetilde{\phi}_u\ .
\end{equation}

Within the relevant regime of $\mathcal{O}_{q_N,p_N}$ (the colored interval of SC$_N$ in \cref{fig:scaledim}, corresponding to \cref{eq: condition for SC_N}), $\mathcal{L}_{ud}$ is an irrelevant perturbation to the free Lagrangian in \cref{eq:SC_fixed point_L}, following the argument in \cite{KFP}. First, note that
\begin{equation}\label{eq:rand-rot}
\partial_x\phi_u =n_z(x) \partial_x\widetilde{\phi}_u+\sqrt{2}\pi\Big[n_+(x) e^{-i\sqrt{2}\widetilde{\phi}_u}+h.c.\Big],
\end{equation}
where $n_\mu(x)$ ($\mu=z,\pm$) are defined by $U(x) \sigma_z U^\dagger(x)=n_z(x)\sigma_z+ n_+(x)\sigma^++n_-(x)\sigma^-$. Since $U(x)$ is defined from a random function $\xi(x)$, we expect $n_\mu(x)$ to be random with $\overline{n_\mu(x)n_\mu(x')}\rightarrow 0$ for $|x-x'|\gg v_u^2/W_{q_N,p_N}$. This provides $\mathcal{L}_{ud}$ with random coefficients $n_\mu(x)$ coupled to scaling dimension $2$ ($>3/2$) operators $\partial_x\widetilde{\phi}_u\partial_x\phi_d$ and $e^{\pm i\sqrt{2}\widetilde{\phi}_u}\partial_x\phi_d$. Thus, $\mathcal{L}_{ud}$ is irrelevant, and the edge theory flows to the stable SC$_N$ fixed point $g=g_N$ with free boson Lagrangian $\mathcal{L}_{\text{SC}_N}$ in \cref{eq:SC_fixed point_L}. Particularly, for the SC$_0$ (KFP) fixed point with $N=0$ (studied in \cite{KFP}), $\phi_d$ and $\widetilde{\phi}_u$ reduce to $\phi_\rho$ and $\widetilde{\phi}_\sigma$ ($\phi_\sigma$ after random SU(2) rotation) by \cref{eq: change_basis_charge_neutral}.

\begin{figure}[h!]
\centering
\begin{tabular}{ l l }
    (a)\\
    \resizebox{\columnwidth}{!}{\includegraphics{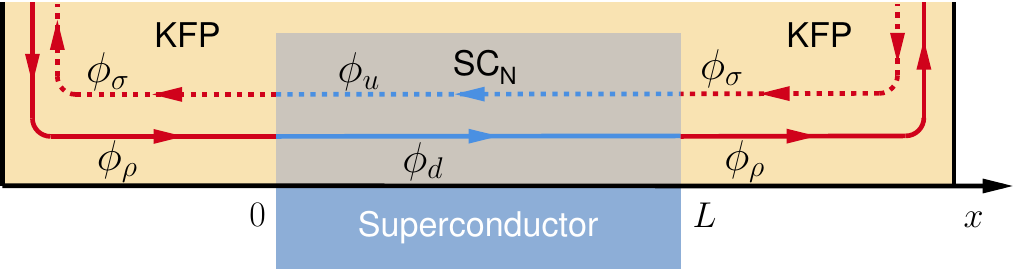}}\\

    (b)\\
    \resizebox{\columnwidth}{!}{\includegraphics{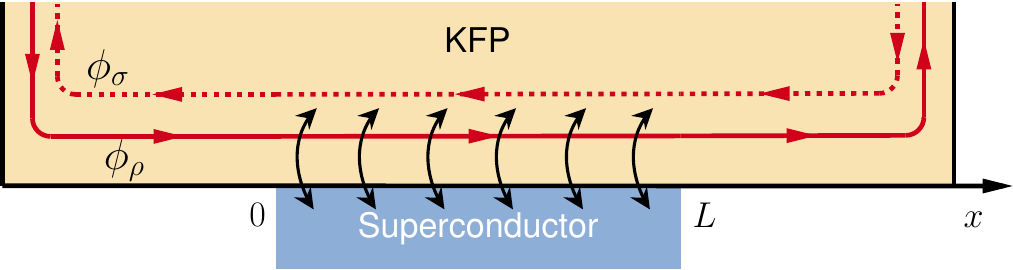}}
\end{tabular}
\caption{
The lower edge of \cref{fig:setup} which constitutes a KFP-SC$_N$-KFP junction, with the SC-proximitized edge interval $0<x<L$ in the SC$_N$ phase, and the rest of the edge in the KFP phase. (a) The $N\neq 0$ case, where the SC proximity is strong and results in a quantized downstream resistance $R_d$ in \cref{eq: Rd}. (b) The $N=0$ case, where the SC proximity is inefficient and leads to a nonlinear $R_d$ in \cref{eq: Rd-non}.
}
    \label{fig:KFP-SC-KFP}
\end{figure}

{\color{blue}\emph{Quantized transport for $N\neq 0$.}} We now show that an edge in the SC$_{N\neq0}$ phase in Fig. \ref{fig:setup} leads to quantized electric transport at zero temperature. Figure \ref{fig:KFP-SC-KFP}(a) shows the lower edge of Fig. \ref{fig:setup}. In the long-edge and zero-temperature limit, we assume the edge in intervals $x<0$ and $x>L$ without SC proximity flows to the KFP fixed point (as consistent with past experiments), and the edge in interval $0<x<L$ with SC proximity flows to the SC$_N$ fixed point. Thus, the lower edge constitutes a KFP-SC$_N$-KFP junction: each interval has a free downstream boson ($\phi_d$ in SC$_N$, and $\phi_\rho$ in KFP), and an upstream boson which is randomly rotated by the relevant disorder ($\phi_u\rightarrow \widetilde{\phi}_u$ in SC$_N$, $\phi_\sigma\rightarrow \widetilde{\phi}_\sigma$ in KFP, by \cref{eq:rand-rot}) and RG flows to free. The electric transport of the lower edge is governed by its charge transmission $t_{eh}=t_n-t_a$ across the SC$_N$ interval, where $t_n$ and $t_a$ are the normal and Andreev (electron-to-hole conversion) transmissions, respectively, with $t_n+t_a=1$.

To calculate $t_{eh}$, we apply a voltage $V(x)$ (defined relative to SC) which is only nonzero within $x<0$, and calculate the current $I(x)$ at $x>L$. By the Kubo formula, $I(x) = \frac{e^2}{2\pi h}\int_0^\infty dt\int_{-\infty}^{\infty} dx' V(x')D(x,x',t)$, where the propagator $D(x,x',t) \equiv -i\frac{4\pi^2}{e^2}\braket{[j(x,t),\rho(x',0)]}$. In the basis $(\phi_\sigma,\phi_\rho)^T$ in \cref{eq: def charge and neutral basis}, the charge density $\rho= -\frac{e}{2\pi} \sqrt{\frac{2}{3}} \partial_x \phi_\rho$ and current $j=  \frac{e}{2\pi} \sqrt{\frac{2}{3}} \partial_t \phi_\rho$ only depend on $\phi_\rho$. For calculating current $I(x)$ at $x<0$ or $x>L$ in the KFP intervals where $\phi_\rho$ propagates freely with velocity $v_\rho$, one has $j=v_\rho\rho$ and thus the propagator can be rewritten as
\begin{equation}\label{eq:initial_condition}
    D(x,x',t) = -\frac{2i}{3}v_\rho\braket{[\partial_x\phi_\rho(x,t),\partial_{x'}\phi_\rho(x',0)]}\ .
\end{equation}

It is sufficient to set $x'<0$, as $V(x')$ is zero elsewhere. Solving the zero-temperature $D(x,x',t)$ is equivalent to solving the equation of motion of boson field $\partial_x\phi_\rho(x,t)$ with initial condition $\partial_x\phi_\rho(x,0)=-\frac{4\pi}{3}v_\rho\partial_x\delta(x-x')$. With $x'<0$, the mode $\phi_\rho$ initially at $x<0$ scatters into $\phi_d$ and $\phi_u$ at $0<x<L$, 
then turns back to $\phi_\rho$ at $x>L$. At the interval boundaries $x=0$ and $x=L$, \cref{eq: change_basis_charge_neutral} imposes a boundary condition $(-v_u\partial_x\phi_u,v_d\partial_x\phi_d)^T|_{L^-/0^+}=\mathcal{R}_N(-v_\sigma\partial_x\phi_\sigma,v_\rho\partial_x\phi_\rho)^T|_{L^+/0^-}$ (see SM \cite{supp}). In the KFP and SC$_N$ intervals, while $\phi_\rho$ and $\phi_d$ are free modes, $\phi_\sigma$ and $\phi_u$ differ from the free modes $\widetilde{\phi}_\sigma$ or $\widetilde{\phi}_u$ by a nonlinear transformation \cref{eq:rand-rot}, making it difficult to solve the boundary condition. However, randomness simplifies the problem. Let us set $\partial_x\phi_u\vert_{L^-} = \partial_x\widetilde{\phi}_u\vert_{L^-}$ by choosing $x_0=L$ in the definition of $U(x)$. By \cref{eq:rand-rot}, the upstream current $\partial_x\phi_u\vert_{0^+}$ at end-point $x=0^+$ is related to the free currents of $\widetilde{\phi}_u$ from $x=L$ by random coefficients $n_{\mu}(0)$. Since $n_{\mu}(0)$ depends on the energy-dependent effective velocities $v_{ij}$ and the possibly dynamically fluctuating SC disorder $\xi(x')$, the averaging over particle energies and time effectively averages $n_{\mu}(0)$ to zero and leads to $\partial_x\phi_u\vert_{0^+}=0$ for large $L$ \cite{supp}. 
Also, the upstream current $\partial_x\phi_\sigma|_{L^+}=0$ at end-point $x=L^+$, since there is no incoming upstream current of $\widetilde{\phi}_\sigma$.
Altogether, the boundary conditions are simplified into
\begin{equation}\label{eq:simple-bc}
\begin{split}
&\left.v_\rho\partial_x\phi_\rho\right|_{0^-}=\left.p_Nv_d\partial_x\phi_d\right|_{0^+}\ ,\\ &\left.v_d\partial_x\phi_d\right|_{L^-}=\left.p_Nv_\rho\partial_x\phi_\rho\right|_{L^+}\ .
\end{split}
\end{equation}
With $\phi_\rho$ and $\phi_d$ being free, \cref{eq:simple-bc} indicates that the scattering from $x<0$ to $x>L$ simply reduces $\partial_x\phi_\rho$ by a factor $1/p_N^2$ (factor $1/p_N$ at each boundary). This solves the zero-temperature $D(x,x',t)$ for $x'<0$ and $x>L$ as $D(x, x', t) = \frac{4\pi}{3p_N^2}\partial_t\delta(x-x'-v_\rho t-L_0)$
with $L_0=(1-\frac{v_\rho}{v_d})L$. The Kubo formula then yields a downstream current $I(x>L)=\frac{2 e^2}{3 p_N^2 h} V(-\infty)=t_{eh}\frac{\nu e^2}{h} V(-\infty)$, giving a \emph{quantized} charge transmission $t_{eh}=1/p_N^2$.

In \cref{fig:setup}, current $I$ flows from lead $1$ into the SC connected to lead 3, while lead 2 is floating. We denote the voltage of lead $j$ ($j=1,2,3$) as $V_j$ (so the SC is at voltage $V_3$), and define $V_{ij}=V_i-V_j$. From the generalized Landauer-B\"uttiker formula \cite{BTK1982, Anantram1996_AndreevScattering, Aharony2008, hu2024resistance} (see SM \cite{supp}), for $N\neq0$, we arrive at a zero-temperature \emph{quantized} downstream resistance:
\begin{equation}\label{eq: Rd}
    R_d \equiv \frac{V_{23}}{I} = \frac{h}{\nu e^2}\frac{t_{eh}}{1-t_{eh}} = \frac{h}{2q^2_N e^2} \ .
\end{equation}
If $N=0$, namely if the proximitized edge stays in the KFP phase (\cref{fig:KFP-SC-KFP}(b)), $q_N=0$ and \cref{eq: Rd} diverges, which gives way to higher order contributions, as discussed below. 

Equation \eqref{eq: Rd} represents a distinctive quantized transport unique to the FQH-SC junction. Intuitively, the quantization occurs because the upstream edge current $\partial_x\phi_u$ averages to zero due to its strong random coupling to the SC, leaving only the free downstream edge current $\partial_x\phi_d$ decoupled from the SC. Notably, \cref{eq: Rd} is distinct from the four-terminal Hall resistance, which remains quantized at $R_H = \frac{3}{2}\frac{h}{e^2}$ irrespective of the phase of the SC proximitized edge \cite{supp}.

{\color{blue}\emph{Higher order nonlinear transport.}} As higher order effects, the edge current can also flow into the SC via all the irrelevant $q\neq0$ Cooper pair tunneling terms $\mathcal{O}_{q,p}$ in \cref{eq:full_L_disorder}, among which $\mathcal{O}_{p_N,3q_N}$ and $\mathcal{O}_{-p_N,-3q_N}$ have the lowest scaling dimension $\Delta_{p_N,3q_N}=3$ and give the leading contribution (see SM \cite{supp}). In a steady current state driven by a nonzero voltage $V_{13}$ between lead $1$ and the SC (lead 3), this adds a tunneling Hamiltonian $H_T=\int_0^L dx [e^{i\frac{2e}{\hbar}q_\text{eff}V_{13}t}\xi_{p_N,3q_N}(x)\mathcal{O}_{p_N,3q_N} +h.c.]$ with certain $q_\text{eff}\neq 0$, and the tunneling current operator $\hat{I}_T= \frac{p_N}{q_\text{eff}V_{13}}\frac{\partial H_T}{\partial t}$. At zero temperature, this tunnels an extra current $I_T=\frac{i}{\hbar}\int_{-\infty}^0 dt\langle\overline{[H_T(t),\hat{I}_T(0)]}\rangle\simeq W_{T} L |V_{13}|^{\alpha}V_{13}$ into the SC, where the exponent $\alpha=2\Delta_{p_N,3q_N}-2=4$, and $W_T\propto |\xi_{p_N,3q_N}|^2$ (see SM \cite{supp}). This yields a correction to the charge transmission $\delta t_{eh}=t_{eh}-1/p_N^2\propto -W_{T} L |V_{13}|^{\alpha}$, and a nonlinear correction to $R_d$ of \cref{eq: Rd} at temperatures $T\ll e|V_{13}|/k_B$ and $|V_{13}|\ll (W_TL)^{-1/\alpha}$, with $\alpha=4$:
\begin{equation}\label{eq: Rd-non}
\begin{split}
&R_d-\frac{h}{2q^2_N e^2}\propto -W_T L |V_{13}|^\alpha\ ,\qquad (N\neq 0)\\
&R_d\propto W_T^{-1} L^{-1} |V_{13}|^{-\alpha}\ .\qquad\qquad\ \ \ (N =0)
\end{split}
\end{equation}
If temperature $T\gg e|V_{13}|/k_B$ and $k_BT/e\ll (W_TL)^{-1/\alpha}$, the $R_d$ correction becomes \cref{eq: Rd-non} with $|V_{13}|$ substituted by $k_BT/e$ (see SM \cite{supp}). Specifically, our result for $N=0$ agrees with an earlier result in \cite{Fisher1994}.

Additionally, dissipation into near-edge SC vortices (or metallic puddles) from the magnetic field and thermal activation may occur and affect the transport \cite{zhao2020interference, kim2022CAR, zhao2023loss, Schiller2023,kurilovich2023disorder,hu2024resistance}. We treat each such vortex as a gapless electron reservoir, to which the edge states has a tunneling term $\mathcal{L}_v=\sum_r\zeta_r(x)f^\dagger \mathcal{K}_{r} +h.c.$, where $\mathcal{K}_{r}=e^{i(r+1)\phi_1+i3r\phi_2}=e^{i\sqrt{\frac{3}{2}}\phi_\rho+i\frac{2r-1}{\sqrt{2}}\phi_\sigma}$ ($r\in\mathbb{Z}$) is a generic edge fermion operator annihilating charge $-e$, and $f^\dag$ creates an electron in the reservoir. Assuming the reservoir has a nearly constant density of states, one finds $\langle f(x,t)f^\dagger(x,0) \rangle\propto t^{-1}$ (see SM \cite{supp}), and $\mathcal{K}_{r}$ is a relevant term only if its scaling dimension $\Delta_{r}<\frac{1}{2}$. Generically, $\Delta_{r}$ is lower bounded by conformal spin $s_r$ as $\Delta_{r}\ge s_r=\frac{|3-(2r-1)^2|}{4}\ge\frac{1}{2}$; at the fixed points SC$_N$, it is further lower bounded by $\Delta_{r}\ge 1$, with equality reachable if and only if $N=0,\pm1$. Thus, the effect of each vortex alone is irrelevant. Treating these vortices as uncorrelated and perturbative, by considering a tunneling Hamiltonian $H_T=\int_0^L dx[e^{i\frac{e}{\hbar}q_\text{eff}V_{13}t}\zeta_r(x)f^\dagger \mathcal{K}_{r}+h.c.]$ with the smallest $\Delta_r$, one arrives at a nonlinear correction to $R_d$ similar to \cref{eq: Rd-non}, with a different exponent $\alpha=2\Delta_r-1$, and $W_T$ proportional to the density of near-edge vortices. For $N=0,\pm1$, this yields $\alpha=1$ which dominates over the aformentioned Cooper pair tunneling ($\alpha=4$). For other $N$, this gives $\alpha\gg 4$ which is negligible. Furthermore, multi-electron vortex tunnelings can be similarly analyzed (see SM \cite{supp}), and the only case they dominate is at $N=\pm2$, which yields $\alpha=3$ as contributed by three-electron tunnelings. For $|N|>2$, all vortex tunnelings yield $\alpha\gg 4$, hence negligible compared to the Cooper pair tunneling ($\alpha=4$).

{\color{blue}\emph{Discussion.}}
We have presented a theory of quantized transport induced by the SC proximity to the paradigmatic $\nu = 2/3$ FQH edge state hosting counter-propagating modes. We show that interactions combined with disordered tunneling into the SC can induce an \textit{infinite} set of stable edge phases termed as SC$_{N}$. Each SC$_N$ edge phase exhibits a topological transport signature in a FQH-SC junction, as summarized by \cref{eq: Rd} for the leading quantized effect and \cref{eq: Rd-non} for nonlinear corrections with a universal scaling. 

Regarding realizations, from \cref{eq: condition for SC_N} and \cref{fig:scaledim}, we see that the $N\ge0$ ($N<0$) phases require a repulsive (attractive) intermode interaction $v_{12}>0$ ($v_{12}<0$). We expect the KFP and SC$_{-1}$ phases to be the most probable in experiments, requiring only a moderate interaction $v_{12}$. The attractive interaction needed for the SC$_{-1}$ phase may be induced by the attractive electron-electron interaction in SC or the electron-phonon coupling enhanced by interfacial effects \cite{lee2014interfacial}. Moreover, SC with zero or low density of magnetic vortices is desired for observing the quantized $R_d$ in \cref{eq: Rd}, for which FCI at zero magnetic field may be advantageous compared to FQH systems in nonzero field.

Our theory straightforwardly applies to the bilayer Halperin-$(112)$ \cite{halperin1983theory} (or spin-unpolarized \cite{Yutushui2024}) $\nu=2/3$ FQH state, which is topologically equivalent to the single-layer $\nu=2/3$ state studied here, with edge boson fields of the two layers $(\phi_1',\phi_2')^T=(-\phi_2,\phi_1+2\phi_2)^T$ linearly related to $(\phi_1,\phi_2)^T$ in \cref{eq:2/3_Lagrangian_clean}. The interaction parameter in \cref{eq: def_g} becomes $g=g_0\frac{(2-\sqrt{3})v_{11}'+(2+\sqrt{3})v_{22}'-2v_{12}'}{(2+\sqrt{3})v_{11}'+(2-\sqrt{3})v_{22}'-2v_{12}'}$ in terms of the bilayer basis velocity matrix $v_{ij}'$ \cite{supp}. While the SC$_0$ (KFP) phase occurs near equal intralayer velocities $v_{11}'\approx v_{22}'$, the SC$_{N\neq 0}$ phases can be realized by increasing $|v_{11}'-v_{22}'|$. Experimental control of edge interactions might be achieved by gate-tuning the dielectric environment \cite{jia2022tuning, wang2023robust}. Besides, the interface between $\nu=1$ and $\nu=1/3$ states also serves as a promising platform for realizing our proposal \cite{cohen2019synthesizing}.

It remains unclear in the regime without relevant disorders $\mathcal{O}_{q,p}$ (uncolored intervals in \cref{fig:scaledim}), whether $R_d$ will be non-universal or will eventually equilibrate to universal values by irrelevant scatterings. Generically, disorders $\mathcal{O}_{q,p}$ with $(q,p)$ not limited to Eq. \eqref{eq:p_q_relation}, even if perturbatively irrelevant, may still impact the edge phases and transport if their bare strengths are large enough. Recent studies in non-SC setups have revealed a crossover between coherent and incoherent edge transport regimes \cite{protopopov2017transport, cohen2019synthesizing, Hashisaka2023, manna2023diagnostics, manna2024shot}, and it will be interesting to explore the SC generalization. Moreover, at moderate temperatures, additional counter-propagating edge modes may arise from edge reconstruction \cite{Meir1994, KunYang2003, WMG2013}, the effect of which calls for future studies. However, both theories and experiments in non-SC setup confirm that these additional modes will be eliminated by relevant backscatterings and give way to the KFP phase at sufficiently low temperatures \cite{WMG2013}, thus we expect the same holds for the SC$_N$ phases. More generally, it would be intriguing to explore other Abelian \cite{KaneFisher1995, Moore1998} and non-Abelian \cite{levin2007,lee2007,lianwang2018} FQH states hosting multiple counter-propagating edge modes, the SC proximity of which may lead to new stable edge phases beyond our current classification.


\begin{acknowledgments}
\emph{Acknowledgments.} We are grateful to Andrei Bernevig, Leonid Glazman, Duncan Haldane, Zhurun Ji, Yanyu Jia, Charles Kane, Philip Kim, Vlad Kurilovich, Yves Kwan, Nicolas Regnault,  Sanfeng Wu and Wucheng Zhang for helpful discussions. This work is supported by the National Science Foundation under award DMR-2141966, and the National Science Foundation through Princeton University’s Materials Research Science and Engineering Center DMR-2011750. Additional support is provided by the Gordon and Betty Moore Foundation through Grant GBMF8685 towards the Princeton theory program. P.M.T. is supported by a postdoctoral research fellowship at the Princeton Center for Theoretical Science and a Croucher Fellowship. H.C. receives additional supports from Bede Liu Fund for Excellence at the Department of Electrical and Computer Engineering of Princeton University.

\end{acknowledgments}

\bibliographystyle{apsrev4-2.bst}
\bibliography{reference}

\clearpage
\newpage

\clearpage
\newpage
\widetext

\begin{center}
\textbf{\large Supplemental Materials for ``Quantized Transport of $\nu = 2/3$ Fractional Quantum Hall Edge \\ with Disordered Superconducting Proximity"}\\
\vspace{0.5cm}
\text{Pok Man Tam, Hao Chen, and Biao Lian}
\end{center}
\maketitle
\onecolumngrid
\setcounter{secnumdepth}{3}
\renewcommand{\theequation}{\thesection.\arabic{equation}}
\renewcommand{\theHequation}{\theHsection.\arabic{equation}}
\renewcommand{\thefigure}{\thesection.\arabic{figure}}  

The supplemental information consists of five sections. In Sec. \ref{sec:vertex operator}, we provide explicit calculation for the scaling dimensions of vertex operators considered in this work, and elaborate on their dependence on the edge interaction parameter. In Sec. \ref{sec: RG}, we briefly review the disorder renormalization group (RG) analysis and clarify on the effect of chemical potential disorder. We also demonstrate the decoupling of up- and down-stream modes at the RG fixed points as $v_{ud}\propto(g-g_N)$. In Sec. \ref{sec: scattering}, we derive the boundary conditions for the scattering problem of the KFP-SC$_N$-KFP junction (see Fig. \ref{fig:KFP-SC-KFP}(a) in the main text), justify the disorder-averaged boundary conditions, and discuss the application of the Landauer-B\"uttiker formula for obtaining the downstream resistance that is shown to be quantized in the main text. As a contrast, we also show that the Hall resistance is insensitive to the SC$_N$ phases. In Sec. \ref{sec: nonlinear}, we detail the calculation for the higher order nonlinear transport effects that arise from either Cooper-pair tunneling or edge-vortex coupling, as summarized in the main text \cref{eq: Rd-non}. Finally, in Sec. \ref{app_sec: vortex model and RG} we present a vortex model as a statistical ensemble of quantum dots, and explain the RG-irrelevance of the edge-vortex coupling. 

\section{Vertex operators}\label{sec:vertex operator}
\setcounter{equation}{0}
\setcounter{figure}{0} 
\subsection{Setup}
We begin with the minimal model for the clean edge of a $\nu=2/3$ fractional quantum Hall (FQH) state, which is described by the following Lagrangian for a two-component Luttinger liquid theory:
\begin{equation}\label{app_eq:2/3_Lagrangian_clean}
    \mathcal{L}_0 = -\frac{1}{4\pi}\sum_{i,j=1}^2 [K_{ij}\partial_t\phi_i \partial_x \phi_j + v_{ij}\partial_x\phi_i \partial_x\phi_j].
\end{equation}
The topological order is determined by the integer-valued $K$-matrix (up to a $SL(2,\mathbb{Z})$-transformation), while interactions between edge modes are captured by the positive-definite symmetric matrix $v_{ij}$. As a brief review of the $K$-matrix formalism for FQH edge physics, the electric charge density fluctuation is given by $\rho(x) = -\frac{e}{2\pi} \bt^T \partial_x \bphi$, and that of current $j(x)=  \frac{e}{2\pi} \bt^T \partial_t \bphi$ (with $-e<0$ the bare electron charge)
, with the boson field $\bphi=(\phi_1,\phi_2)$, and the charge vector $\bt=(1,1)^T$ in this case. The filling factor is given by $\nu=\bt^TK^{-1}\bt$. The bosonic edge modes obey the commutation relation $[\partial_x\phi_i(x), \phi_j(x')] = 2i\pi\delta(x-x')K^{-1}_{ij}$, and from this one can deduce that the vertex operator $e^{i\bl^T\bphi}$ ($\bl=(l_1,l_2)^T\in\mathbb{Z}^2$) creates a local excitation with charge $e\bl^T K^{-1}\bt$ with the exchange statistical angle $\pi \bl^T K^{-1}\bl$. Notice all allowed vertex operators obey $l_i \in \mathbb{Z}$, so as to be compatible with the $2\pi$ compactification of $\phi_i$. For a detailed review of the $K$-matrix formalism, we refer interested readers to Ref. \cite{WenZee1992, wen2004quantum}. 


In the majority of the main text, we are focusing on the single-layer $\nu=2/3$ FQH state, as described by the following diagonal $K$ matrix under a boson basis $\bphi=(\phi_1,\phi_2)$ and charge vector $\bt=(1,1)^T$ \cite{Wen1990composite, MacDonald1991composite, KFP}:
\begin{equation}\label{app_eq:K matrices def}
    K=K_{\frac{2}{3}} =\begin{pmatrix}
    1 & 0 \\ 0 & -3  \end{pmatrix}\ .
\end{equation}
As alternative physical realization, it is known that the above single-layer $\nu=2/3$ FQH state has the same topological order as the bilayer Halperin-(112) state. Via a $SL(2,\mathbb{Z})$ transformation, the bilayer Halperin-(112) state is described by the following $K$-matrix and boson basis $\bphi'=(\phi_1',\phi_2')^T$ and charge vector $\bt'=\bt=(1,1)^T$ \cite{KaneFisher1995, Yutushui2024}:
\begin{equation}\label{app_eq:K matrices def2}
    \begin{split}
    &K'=K_{(112)}=W^{-1}K_{\frac{2}{3}}({W^{-1}})^{T}=\begin{pmatrix}
    1 & \;\;2 \\ 2 & \;\;1 
    \end{pmatrix},\quad \bphi'=W^T \bphi\ , \quad \bt'=W^{-1}\bt=\bt\ , \quad W=\begin{pmatrix}
    0 & 1 \\ -1 & 2  \end{pmatrix}\in SL(2,\mathbb{Z})\ ,\\
    & \bpm v'_{11} & v'_{12} \\ v'_{12} & v'_{22}\epm = W^{-1} \bpm v_{11} & v_{12} \\ v_{12} & v_{22}\epm (W^{-1})^T = \bpm 4v_{11}-4v_{12}+v_{22} & 2v_{11}-v_{12} \\2v_{11}-v_{12} & v_{11}\epm. 
    \end{split}
\end{equation}
Hereafter, we shall use unprimed notations $K$, $\phi_j$, $v_{ij}$ to denote the parameters of the single-layer $\nu=2/3$ FQH state, and primed notations $K'$, $\phi_j'$, $v_{ij}'$ to refer to the parameters of the bilayer Halperin-(112) state, which are related via the $SL(2,\mathbb{Z})$ transformation in \cref{app_eq:K matrices def2}. Most of the discussions will be focused on the single-layer FQH state.

For the single-layer state, an edge electron can be created by $e^{-i\phi_1}$ or $e^{3i\phi_2}$ (which is chiral in nature). For the bilayer Halperin-(112) state, an edge electron can be created by $e^{-3i\phi_1'}$ or $e^{-3i\phi_2'}$ (which is layer-pseudospin-polarized).

To facilitate the calculation of scaling dimension of vertex operators, we shall first recast the Lagrangian in Eq. \eqref{app_eq:2/3_Lagrangian_clean} into a ``diagonal form" with the $K$-matrix and $v$-matrix simultaneously diagonalized, following the treatment detailed in Ref. \cite{KaneFisher1995, Moore1998}. This is done in three steps: (i) First we use an orthogonal matrix $\Lambda^{(1)}$ to diagonalize $K$: $[(\Lambda^{(1)})^T K \Lambda^{(1)}]_{ij} = \lambda_i \delta_{ij}$. (ii) Then we rescale by $\Lambda^{(2)}_{ij}=\delta_{ij}/\sqrt{\abs{\lambda_i}}$, such that $K_{ij} \rightarrow [(\Lambda^{(1)}\Lambda^{(2)})^T K \Lambda^{(1)}\Lambda^{(2)}]_{ij}= \sgn{(\lambda_i)}\delta_{ij}\equiv \eta_{ij}$ becomes a signature matrix, and $v \rightarrow \bar{v} = [\Lambda^{(1)}\Lambda^{(2)}]^T v \Lambda^{(1)}\Lambda^{(2)}$. (iii) Finally we diagonalize $\bar{v}$ with a pseudo-orthogonal matrix $\Lambda^{(3)}$, such that $(\Lambda^{(3)})^T \eta \Lambda^{(3)}=\eta$ and $[(\Lambda^{(3)})^T \bar{v} \Lambda^{(3)}]_{ij}=\tilde{v}_i \delta_{ij}$. To sum up, with $\phi_i  = \sum_j[\Lambda^{(1)}\Lambda^{(2)}\Lambda^{(3)}]_{ij} \tilde{\phi}_j$, we arrive at the following two decoupled chiral Luttinger liquids:
\begin{equation}\label{app_eq:2/3_Lagrangian_clean_diagonal}
    \mathcal{L}_0 = -\frac{1}{4\pi}\sum_{i=1}^2 [\sgn(\lambda_i)\partial_t\tilde{\phi}_i \partial_x \tilde{\phi}_i + \tilde{v}_{i}\partial_x\tilde{\phi}_i \partial_x\tilde{\phi}_i]. 
\end{equation}
In this theory, the vertex operator $e^{i\bl^T\tilde{\bphi}}$ simply has scaling dimension $(\bl^T\bl)/2$. In other words, the vertex operator $e^{i\bl^T\bphi}$ has scaling dimension
\begin{equation}\label{app_eq: scaling_general}
    \Delta_\bl = \frac{1}{2}\bl^T\Lambda^{(1)}\Lambda^{(2)}\Lambda^{(3)}[\Lambda^{(1)}\Lambda^{(2)}\Lambda^{(3)}]^T\bl,
\end{equation}
which clearly depends on $v_{ij}$-matrix via $\Lambda^{(3)}$. On the other hand, the conformal spin of the vertex operator $e^{i\bl^T\bphi}$ is
\begin{equation}\label{app_eq: spin_general}
    s_\bl= \frac{1}{2}\bl^T\Lambda^{(1)}\Lambda^{(2)}\Lambda^{(3)}\eta(\Lambda^{(3)})^T(\Lambda^{(2)})^T(\Lambda^{(1)})^T\bl = \frac{1}{2} \bl^TK^{-1}\bl.
\end{equation}
where we have used $\Lambda^{(3)}\eta(\Lambda^{(3)})^T=\eta$\;\footnote{For a pseudo-orthogonal matrix $\Lambda$ with respect to a signature matrix $\eta$, i.e., $\Lambda^T\eta\Lambda=\eta$, we have $\Lambda^T=\eta\Lambda^{-1}\eta$, and thus $\Lambda\eta\Lambda^T = \Lambda\eta\eta\Lambda^{-1}\eta = \eta$. }. Hence, the conformal spin $s_\bl$ is independent of $v_{ij}$ and controlled only by the topological order (via $K$).

The above discussion obviously generalizes to complex edge structures with more than two components, but for this specific work we shall focus on the two FQH states with $K$ and $K'$ matrices defined by \cref{app_eq:K matrices def,app_eq:K matrices def2}. With $\eta = \bpm 1 &0\\0 &-1 \epm$, we can parametrize $\Lambda^{(3)}=\bpm \cosh\theta & -\sinh\theta \\ -\sinh\theta &\cosh\theta\epm$. To make sure $\Lambda^{(3)}$ diagonalizes $\bar{v}$, we require $\tanh(2\theta) = 2\bar{v}_{12}/(\bar{v}_{11}+\bar{v}_{22})$, which gives the single interaction parameter that determines the scaling dimension of any given vertex operator. More explicitly, for matrix $K$ (or $K'$) of the single-layer (bilayer) state, we find
\begin{subequations}\label{app_eq: Lambda and g_def}
\begin{align}
    &\text{For single-layer }  \nu=2/3: &&\Lambda^{(1)}\Lambda^{(2)}= \bpm 1 & 0 \\ 0 & \frac{1}{\sqrt{3}}\epm \implies \tanh(2\theta) = \frac{6v_{12}}{\sqrt{3}(3v_{11}+v_{22})},\\
    &\text{For bilayer Halperin (112)}: &&\Lambda^{(1)}\Lambda^{(2)}= \bpm \frac{1}{\sqrt{6}} & -\frac{1}{\sqrt{2}} \\ \frac{1}{\sqrt{6}} & \frac{1}{\sqrt{2}}\epm \implies \tanh(2\theta') = \frac{\sqrt{3}(v'_{22}-v'_{11})}{2(v'_{11}+v'_{22}-v'_{12})},
\end{align}
\end{subequations}
In the main text, see Eq. \eqref{eq: def_g},  we have found it convenient to introduce the interaction parameter $g$ as a function of the velocity matrix $v_{ij}$ of the single-layer FQH state:
\begin{equation}\label{app_eq: Lambda and g_def2}
    g \equiv e^{4\theta} = \frac{1+\tanh{(2\theta)}}{1-\tanh(2\theta)} = \frac{3v_{11}+v_{22}+2\sqrt{3}v_{12}}{3v_{11}+v_{22}-2\sqrt{3}v_{12}},
\end{equation}
which will be shown to acquire a simple expression at the RG fixed point. Using Eq. \eqref{app_eq:K matrices def2}, when transformed into the the bilayer FQH state with velocity parameters $v_{ij}'$, we can rewrite $g$ as
\begin{equation}
    g= g_0 \frac{(2-\sqrt{3})v'_{11}+(2+\sqrt{3})v'_{22}-2v'_{12}}{(2+\sqrt{3})v'_{11}+(2-\sqrt{3})v'_{22}-2v'_{12}},\quad g_0 = (2+\sqrt{3})^2,
\end{equation}
which gives an explicit relation between the interaction parameter controlling scaling dimensions and the microscopic velocity matrix elements in the bilayer realization. This last expression is used in the Discussion of the main text.

\subsection{Scaling dimension}\label{app_sec: basics of scaling dimension}
The scaling dimensions $\Delta_{q,p}$ of the vertex operators $\mathcal{O}_{q,p}= e^{i[q\sqrt{6}\phi_\rho+p\sqrt{2}\phi_\sigma]} = e^{i[(3q+p)\phi_1+3(q+p)\phi_2]}$, evaluated with respect to the clean edge theory Eq. \eqref{app_eq:2/3_Lagrangian_clean}, plays an important role in the RG analysis presented in our main text. Here we calculate how it depends on the interaction parameter $g$. For completeness, we first review some basics of scaling dimensions and conformal spins. Experienced readers can directly skip to the discussion around Eq. \eqref{app_eq: Delta vs g}.

\subsubsection{Overview of the basics}

For the simplest case with a free chiral boson $\phi$ of velocity $v$ obeying canonical commutation relation $[\phi(x),\phi(x')] = i\pi\sgn{(x-x')}$, or equivalently with Lagrangian $-\frac{1}{4\pi}\left[(\partial_t\phi)(\partial_x\phi) + v(\partial_x\phi)^2\right]$, scale invariance of the theory ensures that the two-point correlator of a vertex operator $e^{i\alpha\phi}$ takes the form \cite{bigyellow, appliedCFT,davidtong_string,fradkin2013field}
\begin{equation}
    \langle e^{i\alpha\phi(x,t)} e^{-i\alpha\phi(0,0)}\rangle \propto \frac{1}{(x- vt)^{2h}},
\end{equation}
where the number $h$ is defined as the conformal dimension of $e^{i\alpha\phi}$. The meaning of $h$ is that under rescaling $(x,t)\rightarrow (bx,bt)$, we have $e^{i\alpha\phi(bx,bt)} = b^{-h}e^{i\alpha\phi(x,t)}$. In a free chiral boson theory, $\langle e^{i\alpha\phi(x,t)} e^{-i\alpha\phi(0,0)}\rangle$ is evaluated using the two-point function $\braket{\phi(x,t)\phi(0,0)} = -\log(\frac{x-vt}{a_0})$ (with $a_0$ some short-distance cutoff) 
via Wick's theorem \cite{bigyellow, appliedCFT,davidtong_string,fradkin2013field}:
\begin{equation}
    \langle e^{i\alpha\phi(x,t)} e^{-i\alpha\phi(0,0)}\rangle = e^{\alpha^2\langle\phi(x,t)\phi(0,0)\rangle} \propto \frac{1}{(x- vt)^{\alpha^2}}.
\end{equation}
Thus $h=\frac{\alpha^2}{2}$ for a free chiral boson. 

In a theory consists of multiple free chiral bosons $\phi^c_i$ ($i=1,\dots,n_c$), each with a velocity $v_i>0$, as well as free anti-chiral bosons $\phi^a_j$ ($j = 1,\dots,n_a$) with velocity $-\bar{v}_j<0$, in the absence of inter-mode couplings, the two-point correlator of a vertex operator $\mathcal{O}=e^{i\sum_{i}\alpha_i^c\phi^c_i + i\sum_j \alpha_j^a\phi^a_j}$ has the following form
\begin{equation}
    \left\langle \mathcal{O}(x,t)\mathcal{O}^\dagger(0,0)\right\rangle \propto \prod_{i=1}^{n^c}\frac{1}{(x- v_it)^{2h_i}}\prod_{j=1}^{n^a}\frac{1}{(x+ \bar{v}_jt)^{2\bar{h}_j}} = \prod_{i=1}^{n^c}\frac{1}{(x- v_it)^{(\alpha_i^c)^2}}\prod_{j=1}^{n^a}\frac{1}{(x+ \bar{v}_jt)^{(\alpha_j^a)^2}},
\end{equation}
where $h_i=(\alpha_i^c)^2/2$\;($\bar{h}_j=(\alpha_j^a)^2/2$) is the conformal dimension of this vertex operator in mode $\phi_i^c$ ($\phi_j^a$). The scaling dimension of $\mathcal{O}$ is defined as $\Delta=\sum_{i}h_i+\sum_j\bar{h}_j$, while the conformal spin of $\mathcal{O}$ is defined as $s=\left|\sum_{i}h_i-\sum_j\bar{h}_j\right|$. 

When there are couplings between different modes, i.e., there are terms $v_{ij}\partial_x\phi_i\partial_x\phi_j$ in the Lagrangian where the $v_{ij}$-matrix has nonzero off-diagonal elements, one can no longer directly evaluate the two-point correlators. However, one can always do a change of basis (like $\phi_i  = \sum_j[\Lambda^{(1)}\Lambda^{(2)}\Lambda^{(3)}]_{ij} \tilde{\phi}_j$ in the previous subsection) that diagonalizes the $v_{ij}$-matrix while keeping the canonical commutation relations. Then, in this new basis, the vertex operator $\mathcal{O}$ is expressed in terms of decoupled free bosons $\mathcal{O}=e^{i\sum_{i}\widetilde{\alpha}_i^c\widetilde{\phi}^c_i + i\sum_j\widetilde{\alpha}_j^a\widetilde{\phi}^a_j}$, and the scaling dimension and conformal spin of $\mathcal{O}$ can be calculated as before. This gives rise to \eqref{app_eq: scaling_general} and \eqref{app_eq: spin_general}. A matter of fact is that $s$ is a topologically robust quantity, being insensitive to interaction $v_{ij}$, thus one can always easily evaluate $s$ in any basis by neglecting the interaction. Moreover, since $\sum_{i}h_i\geq0,\ \sum_j\bar{h}_j\geq 0$, $s$ is a lower bound of $\Delta$. 

\subsubsection{Disorder vertex operator}

Now, going back to our FQH edge theory, let us consider the scaling dimension of the vertex operator $\mathcal{O}_{q,p}= e^{i[q\sqrt{6}\phi_\rho+p\sqrt{2}\phi_\sigma]} = e^{i[(3q+p)\phi_1+3(q+p)\phi_2]}$, introduced in main text Eq. \eqref{eq:full_L_disorder} to represent disorder effects. For concreteness, we adopt the single-layer FQH state boson basis $(\phi_1, \phi_2)$ as done in the main text. It can be easily checked that $\mathcal{O}_{q,p}$ annihilates electric charge $-2qe$ and has self-statistical angle $2\pi(3q^2-p^2)$, and thus for integers $(q,p)\in \mathbb{Z}^2$, such a parametrization gives the most general bosonic even-charged operator as allowed by the SC proximity. Following Eq. \eqref{app_eq: scaling_general} and Eq. \eqref{app_eq: Lambda and g_def}, we have:
\begin{equation}\label{app_eq: Delta vs g}
\begin{aligned}
    \Delta_{q,p} =&  2 (p^2 + 3 p q + 3 q^2)\cosh(2\theta) - \sqrt{3} (p + q) (p + 3 q) \sinh(2\theta)\\
    =& \frac{1}{\sqrt{g}}\left[(g+1)(p^2+3pq+3q^2)-\frac{\sqrt{3}}{2}(g-1)(p^2+4pq+3q^2)\right]\\
    =& \frac{1}{2}\sqrt{\frac{g}{g_0}}(p-\sqrt{3}q)^2 + \frac{1}{2}\sqrt{\frac{g_0}{g}}(p+\sqrt{3}q)^2,
\end{aligned}
\end{equation}
where $g_0=(2+\sqrt{3})^2$. 
The main text Figure \ref{fig:scaledim} is plotted accordingly using Eq. \eqref{app_eq: Delta vs g}, for those operators with the lowest attainable scaling dimension and hence can possibly induce an RG-relevant disorder effect. It is apparent that $\Delta_{q,p}$ is invariant under the transformation $q\rightarrow -q,\ \frac{g}{g_0}\rightarrow\frac{g_0}{g}$, giving rise to the reflection symmetry about $g=g_0$ in the main text Figure \ref{fig:scaledim}. 

Upon varying the interaction parameter $g$, it is straightforward to see that the minimal scaling dimension attainable is $\Delta^{\min}_{q,p} = \abs{3q^2-p^2}$. This simple result can be alternatively obtained by noticing that the conformal spin of $\mathcal{O}_{q,p}$ is $\abs{h-\bar{h}}=\abs{3q^2-p^2}$ (easily obtained by tuning to the point where $\phi_\rho$ and $\phi_\sigma$ decouple, and noticing that the spin is a topologically robust quantity insensitive to varying interaction). As the scaling dimension is expressed as $(h+\bar{h})$, the conformal spin provides a strict lower bound to the scaling dimension. The minimal scaling dimension of $\mathcal{O}_{q,p}$ is attained at $g=g_{q,p}$ with the following expression:
\begin{equation}\label{app_eq: g*}
    g_{q,p} = g_0\frac{(p+\sqrt{3}q)^2}{(p-\sqrt{3}q)^2}.
\end{equation}



\subsection{SC$_N$ phase and its fixed point}
As discussed in the main text, and further explained in the next section, the disorder effect from $\mathcal{O}_{q,p}$ is RG-relevant when its scaling dimension $\Delta_{q,p}<3/2$, which drives an RG flow for $g \rightarrow g_{q,p}$. We thus focus on the combination $(q,p)\in\mathbb{Z}^2$ that satisfies $\Delta^{\min}_{q,p} = \abs{3q^2-p^2}=1$. From elementary number theory (using the fact that a perfect square equals $0$ or $1$ mod 4), we see that this requirement implies $p^2-3q^2=1$, which is a case of the so-called Pell-Fermat equation \cite{lenstra2002solving}. The integral solutions (we focus on $p\geq0$ to avoid double-counting the operators and their hermitian conjugates) are labeled by an integer $N$:
\begin{equation}\label{app_eq: q and p}
    q_{N} = \frac{(2+\sqrt{3})^N -(2-\sqrt{3})^N}{2\sqrt{3}},\;\; p_N = \frac{(2+\sqrt{3})^N +(2-\sqrt{3})^N}{2}.
\end{equation}
This set of solutions is simply generated from $p_N+q_N\sqrt{3} = (p_1+q_1\sqrt{3})^N$, given a fundamental solution $(q_1,p_1)=(1,2)$. Substituting Eq. \eqref{app_eq: q and p} into Eq. \eqref{app_eq: g*}, we obtain
\begin{equation}\label{app_eq: gfixed}
    g_N \equiv g_{q_N, p_N} = (2+\sqrt{3})^{4N+2} = g_0^{2N+1},
\end{equation}
which gives the interaction parameter at the SC$_N$ fixed point stated in the main text. This explains why the SC$_N$ fixed points are equally spaced in the scale of $\ln g$, as shown in Fig. \ref{fig:scaledim}.

Let us remark on one more noticeable feature in Fig. \ref{fig:scaledim}: each of the SC$_N$ phase, defined for $\Delta_{q_N, p_N}< \frac{3}{2}$, occupies the same size of the parameter space on the scale of $\ln g$. To see this analytically, notice that from Eq. \eqref{app_eq: Delta vs g} and solving $\Delta_{q_N,p_N}=\frac{3}{2}$, we obtain the boundary of the SC$_N$ phase at $\theta=\theta_N^{(\pm)}$ (or at $g=g_N^{(\pm)}$), with
\begin{equation}\label{app_eq: phase boundary}
    \tanh(2\theta_N^{(\pm)}) = \frac{4\sqrt{3}q_{2N+1}p_{2N+1}\pm3\sqrt{5}}{12q^2_{2N+1}+9}\Longleftrightarrow g^{(\pm)}_N = \frac{7\pm3\sqrt{5}+2(2+\sqrt{3})^{4N+2}}{7\mp 3\sqrt{5}+2(2-\sqrt{3})^{4N+2}}
\end{equation}
Defining $f=\frac{7+3\sqrt{5}}{2}$, $g_0=(2+\sqrt{3})^2$, we have $g^{(\pm)}_N = \frac{f^{\pm 1} +g_0^{2N+1}}{f^{\mp1}+g_0^{-2N-1}}=g_Nf^{\pm 1}$. Thus, $\ln g^{(+)}_N - \ln g^{(-)}_N = 2\ln f$, which is indeed independent of $N$.

Finally, using Eq. \eqref{app_eq: phase boundary} and noticing that $2q_N p_N = q_{2N}$, we obtain the following condition (c.f. Eq. \eqref{eq: condition for SC_N} in the maint text) for realizing the SC$_N$ phase upon varying the edge-mode interactions:
\begin{equation}
    B^{-}_N<\frac{v_{12}}{3v_{11}+v_{22}}<B^{+}_N, \quad\quad \text{with}\;\;B^{\pm}_N \equiv \frac{2q_{4N+2}\pm \sqrt{15}}{24 q^2_{2N+1}+18}.
\end{equation}

\begin{figure}
    \centering
    \resizebox{0.7\columnwidth}{!}{\includegraphics[]{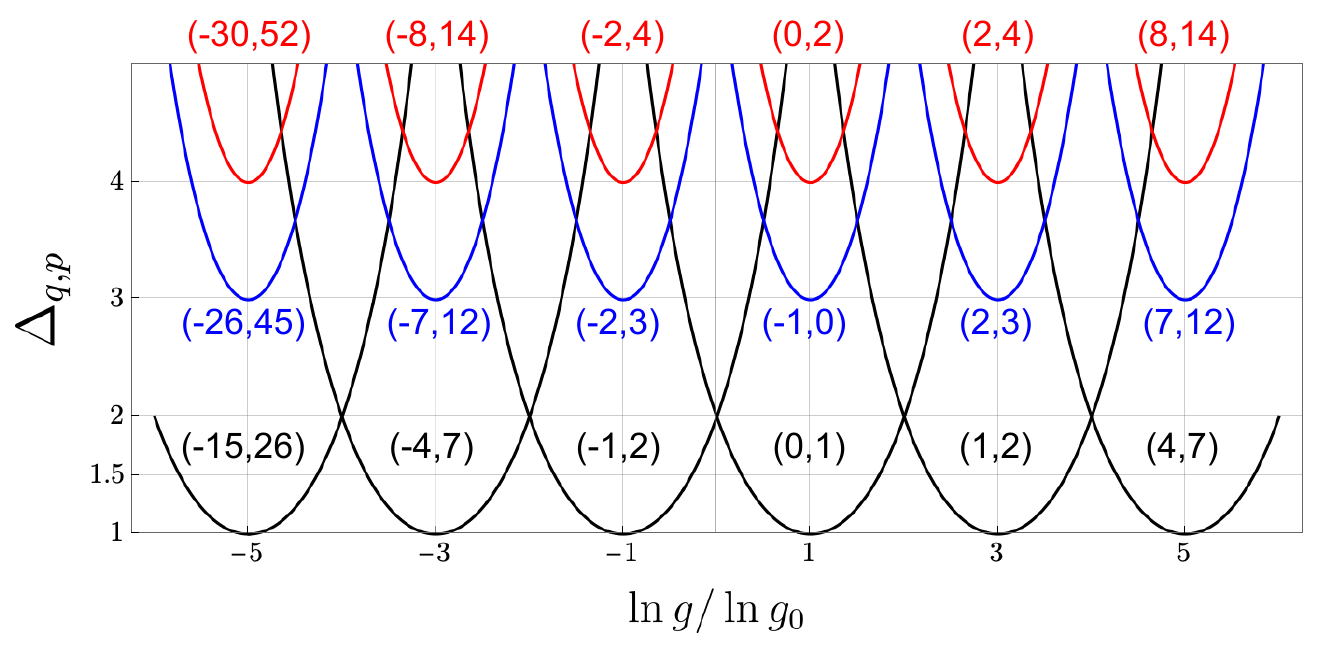}}
    \caption{Scaling dimension $\Delta_{q,p}$ versus the edge-mode interaction parameter $g$. Each curve is labeled by its $(q,p)$, and we take $p>0$ to avoid redundancies. A black curve corresponds to $\mathcal{O}_{\pm q_N, \pm p_N}$, which is the possibly relevant disorder operator that induces the SC$_N$ fixed point. A blue curve corresponds to $\mathcal{O}_{\pm p_N, \pm 3q_N}$, which is the least irrelevant operator that dominates the nonlinear transport at the SC$_N$ fixed point (assuming no vortex tunneling). Red curves are shown for reference, which correspond to the next least irrelevant operator $\mathcal{O}_{\pm 2q_N, \pm 2p_N}$.
    }
    \label{supp_fig:scaledim}
\end{figure}

\subsection{Cooper-pair tunneling operators}\label{app_sec: Cooper-pair scaling dimension}
Towards the end of the main text (around \cref{eq: Rd-non}), we have discussed higher order nonlinear transport effect that arises from perturbative random tunneling of Cooper pairs (or even-charged electron clusters in general). This effect is associated to the RG-irrelevant SC-induced disorder terms $\mathcal{O}_{q,p}$. In the main text we have used the fact that the least irrelevant scaling dimension is $3$ at any SC$_N$ fixed point. Here we provide the explanation. 

Using the fixed point boson basis in the SC$_N$ phase, see Eq. \eqref{eq: change_basis_charge_neutral}, we have $\mathcal{O}_{q,p} \equiv e^{iq\sqrt{6}\phi_\rho +ip\sqrt{2}\phi_\sigma} = e^{i[\sqrt{2}(p_Np-3q_Nq)\phi_u+\sqrt{6}(p_Nq-q_Np)\phi_d]}$, with $q$ and $p$ being integers. Hence, at the SC$_N$ fixed point, the scaling dimension of this operator is 
\begin{equation}\label{sm-DNqp}
    \Delta^{(N)}_{q,p} = (p_Np-3q_Nq)^2+3(p_Nq-q_Np)^2 \quad \in \mathbb{N}.
\end{equation}
Below we will always exclude the trivial case with $q=p=0$ (for which $\mathcal{O}_{q,p}$ is identity). To minimize $\Delta^{(N)}_{q,p}$ we consider two cases for which one of the two terms in \cref{sm-DNqp} is first minimized: (I) Firstly let us assume $p_Nq-q_Np = 0 \implies q=q_Np/p_N$, in which case $\Delta^{(N)}_{q,p} = (p/p_N)^2$ upon using the fixed-point condition: $p_N^2-3q_N^2=1$. Excluding the trivial case where $q=p=0$, we see that the minimal scaling dimension is achieved at $(q,p)=\pm(q_N, p_N)$. The corresponding operator $\mathcal{O}_{\pm q_N, \pm p_N}$ is precisely the relevant disorder operator inducing the SC$_N$ fixed point. The next smallest dimension is then achieved at $(q,p)=\pm(2q_N, 2p_N)$, with $\Delta^{(N)}_{\pm 2q_N, \pm 2p_N} = 4$.  (II) Secondly let us assume $p_Np-3q_Nq=0$, in which case $p_Nq-q_Np$ is obviously non-zero. In this case, the smallest scaling dimension is reached at $(q,p) = \pm(p_N, 3q_N)$ such that $\Delta^{(N)}_{\pm p_N, \pm 3q_N} = 3$. Lastly, if both terms in \cref{sm-DNqp} are nonzero, it is obvious that $\Delta^{(N)}_{q,p}\ge1+3=4$. Summing up, the least irrelevant operator at any SC$_N$ fixed point is $\mathcal{O}_{\pm p_N, \pm 3q_N}$ with dimension $3$, followed by $\mathcal{O}_{\pm 2q_N, \pm 2p_N}$ with dimension 4. Based on Eq. \eqref{app_eq: Delta vs g}, Fig. \ref{supp_fig:scaledim} shows the variation of the scaling dimension of the possibly relevant, as well as the least irrelevant, disorder operators as a function of the interaction parameter $g$. 

\subsection{Vortex tunneling operators}\label{app_sec: scaling dimension: vortex tunneling}
Another contribution to the nonlinear transport, as discussed in the main text, comes from the tunneling of individual electrons between the edge states and the near-edge vortices (which can be viewed as normal metallic quantum dots) inside the SC lead. We will provide an elaborate discussion on this effect in a later section of the supplementary materials. In general multiple electrons can tunnel all at once into a vortex (to be modeled as a quantum dot with many flavors of electrons and having a constant density of states, see Sec. \ref{app_sec: vortex model and RG} for more details), and as discussed in the main text and with more details provided in Sec. \ref{app_sec: nonlinear current: vortex tunneling}, the tunneling exponent $\alpha$ in Eq. \eqref{eq: Rd-non} from charge $Qe$ tunnelings takes the form $\alpha=2\Delta+\abs{Q}-2$, with $\Delta$ the scaling dimension of the least irrelevant charge-$Qe$ vertex operator $\mathcal{O}$ tunneling $Q$ electrons from the edge states into the vortex. 
Recall that the irrelevant Cooper-pair tunneling into the SC lead already gives $\alpha=4$, while tunneling them into the vortex would only lead to a higher $\alpha$ (hence ignored). As far as the vortex tunneling is concerned, for the case of even charge $Q\in2\mathbb{Z}$, the tunneling is to the leading order given by the vertex operator $\mathcal{O}_{q_N,p_N}$ as discussed above, which has scaling dimension $\Delta=1$ at the SC$_N$ fixed point and thus $\alpha=2\abs{q_N}$. The only non-trivial case here is for $N=\pm1$, which gives $\alpha=2$ for the vortex tunneling exponent. Below we see that a smaller $\alpha$ results from tunneling a single electron for $N=\pm1$. 

Let us now focus on odd charge $Q\in 2\mathbb{Z}+1$ vertex operators to look for a possibly smaller $\alpha$. The fermionic nature of such vertex operators dictates $\Delta\geq 1/2$ (since their conformal spins are at least $1/2$), thus $\alpha=2\Delta+Q-2 \geq Q-1$. To search for exponents $\alpha<4$, we only need to focus on odd $Q=1,3$. For odd $Q$, the edge electron cluster operator generally takes the form (with $r\in\mathbb{Z}$)
\begin{equation}\label{app_eq: odd-charge cluster operator}
    \mathcal{K}_{Q,r} = e^{iQ\sqrt{\frac{3}{2}}\phi_\rho +i\frac{2r-1}{\sqrt{2}}\phi_\sigma}
\end{equation}
and its scaling dimension at the SC$_N$ fixed points is
\begin{equation}
    \Delta_{\mathcal{K}_{Q,r}}= \left[\frac{3}{4}Q^2+\frac{1}{4}(2r-1)^2\right](6q_N^2+1)-3Q(2r-1)q_N\sqrt{3q_N^2+1} \quad \in\mathbb{Z}.
\end{equation}
$\Delta_{\mathcal{K}_{Q,r}}$ has to be an integer because $\left[\frac{3}{4}Q^2+\frac{1}{4}(2r-1)^2\right]\in\mathbb{Z}$ for odd $Q$ and integer $r$. Thus $\Delta_{\mathcal{K}_{Q,r}}\geq 1$. It is further minimized at $r=r^*=\lfloor{\frac{3Qq_N\sqrt{3q_N^2+1}}{6q_N^2+1}+\frac{1}{2}}\rceil$, where $\lfloor{...\rceil}$ means taking the nearest integer. 

(i) For $Q=1$, we have $r^*=0$ for $N\leq 0$ and $r^*=1$ for $N\geq 0$, thus
\begin{equation}
    \min\Delta_{\mathcal{K}_{1,r}}= 1+6q_N^2-3\abs{q_N}\sqrt{3q_N^2+1}\geq 1.
\end{equation}
For $\abs{N}>1$ we have $\min\Delta_{\mathcal{K}_{1,r}} \geq 13$, which is highly irrelevant. The above lower bound is only saturated at $N=0,\pm1$, which gives $\alpha=1$ for the vortex tunneling exponent.

(ii) For $Q=3$ and $N=0$, we have $r^*=0,1$, which gives $\min\Delta_{\mathcal{K}_{3,r}}=7$, and hence highly irrelevant. For $Q=3$ and $N>0$ we have $r^*=3$, while for $Q=3$ and $N<0$ we have $r^*=-2$, thus
\begin{equation}
    \min\Delta_{\mathcal{K}_{3,r}}= 13(6q_N^2+1)-45\abs{q_N}\sqrt{3q_N^2+1} \geq 1.
\end{equation}
For $\abs{N}>2$, $\min\Delta_{\mathcal{K}_{3,r}} \geq 13$, which is again highly suppressed. The lower bound is only saturated at $N=\pm1, \pm2$, which gives $\alpha=3$.

Summing up, the above scaling dimension analysis suggests that (i) for $N=0,\pm 1$, the smallest vortex tunneling exponent is $\alpha=1$ via a charge-$e$ tunneling; (ii) for $N=\pm2$, the smallest vortex tunneling exponent is $\alpha=3$ via a charge-$3e$ tunneling; (iii) for $\abs{N}>2$, vortex tunneling always leads to $\alpha > 4$, i.e., highly irrelevant when compared to the Cooper-pair tunneling into the SC lead.

\section{Disorder renormalization group analysis}\label{sec: RG}
\setcounter{equation}{0}
\setcounter{figure}{0} 
\subsection{Criterion for RG relevance}
Here we review the criterion for a spatially random perturbation to be relevant under RG. We closely follow the analysis in Ref. \cite{Kane_edgetransport, GiamarchiSchulz1988}, and work in the imaginary-time $\tau=it$ here ($t$ is real time). In the presence of a random perturbation due to $\mathcal{O}$, the total action for the $(1+1)$-dimensional edge takes the form
\begin{equation}
    S= S_0 +\int \diff{x}\diff{\tau}\;\xi(x) \mathcal{O}(x,\tau) + h.c.
\end{equation}
with $S_0$ derived from Eq. \eqref{app_eq:2/3_Lagrangian_clean} and $\xi(x)$ is modeled to satisfy $\overline{\xi(x)\xi^*(x')}=W\delta(x-x')$ with a zero-mean Gaussian distribution. The overline represents disorder-averaging. The partition function $\mathcal{Z} = \int[\mathcal{D}\phi]e^{-S}$ is then averaged to be
\begin{equation}
\begin{split}
    \overline{\mathcal{Z}}=\int [\mathcal{D\phi}]\;e^{-S_0}\Big[1+\frac{1}{2}\int\diff{x}\diff{\tau}\diff{x'}\diff{\tau'} \;\Big(\overline{\xi(x)\xi^*(x')}\mathcal{O}(x,\tau)\mathcal{O}^\dagger(x',\tau') + \overline{\xi^*(x)\xi(x')}\mathcal{O}^\dagger(x,\tau)\mathcal{O}(x',\tau')\Big)+ ...\Big].
\end{split}
\end{equation}
Thus to the leading order in the disorder strength $W$, the effective action upon disorder-averaging ($\overline{\mathcal{Z}}\equiv \int[\mathcal{D}\phi]e^{-S_{\text{eff}}}$)  takes the form
\begin{equation}
    S_{\text{eff}} = S_0 - W\int \diff{x} \diff{\tau}\diff{\tau'} \mathcal{O}^\dagger(x,\tau)\mathcal{O}(x,\tau'). 
\end{equation}
Under the RG transformation consisting of (1) integrating out high-energy modes in the momentum shell $b\Lambda<k<\Lambda$, and (2) rescaling space and time by $x\rightarrow b^{-1}x$ and $\tau\rightarrow b^{-1}\tau$, the perturbation in the effective action transforms as $-b^{-3+2\Delta}W\int \diff{x} \diff{\tau}\diff{\tau'} \mathcal{O}^\dagger(x,\tau)\mathcal{O}(x,\tau')$, and the RG equation for $W$ reads
\begin{equation}
    \frac{\diff{W}}{\diff{\lambda}} = (3-2\Delta)W
\end{equation}
with $b\equiv e^{-\lambda}$ and $\Delta$ the scaling dimension of $\mathcal{O}$. 
Thus we arrive at the criterion used in the main text: the disorder potential due to $\mathcal{O}$ is relevant (irrelevant) if its scaling dimension $\Delta$ is less (greater) than $3/2$.

\subsection{Chemical potential disorder}
While we have considered in the main text the effect of random disorder coming from vertex operators (which describe either scattering or pairing among edge channels), one may wonder if there are disorder effects due to randomness in the chemical potential. Here we explain why the effect of chemical potential disorder is unimportant, except it can actually induce further randomness in the vertex operator terms that we have been discussing. \\

Chemical potential disorder enters into the Lagrangian in Eq. \eqref{app_eq:2/3_Lagrangian_clean} as $\frac{1}{2\pi}\sum_i \mu_i(x)\partial_x\phi_i$. Notice that such a random chemical potential term can be eliminated via the following shift transformation
\begin{equation}\label{eq:transformation to remove chemical disorder}
    \phi_i(x) \rightarrow \phi_i(x) + \int_{-\infty}^x \diff{x'} \sum_j v^{-1}_{ij} \mu_j(x').
\end{equation}
Under the above transformation: 
\begin{align*}
K_{ij}\partial_t\phi_i \partial_x \phi_j &\rightarrow K_{ij}\partial_t\phi_i [\partial_x \phi_j+\sum_k v^{-1}_{jk}\mu_k(x)] = K_{ij}\partial_t\phi_i \partial_x \phi_j + \text{total time-derivative}\\
 v_{ij}\partial_x\phi_i \partial_x\phi_j &\rightarrow v_{ij} [\partial_x \phi_i+\sum_l v^{-1}_{il}\mu_l(x)] [\partial_x \phi_j+\sum_k v^{-1}_{jk}\mu_k(x)] =  v_{ij}\partial_x\phi_i \partial_x\phi_j + 2\mu_i(x)\partial_x\phi_i + \text{constant}\\
 \mu_i(x)\partial_x\phi_i &\rightarrow \mu_i(x)\partial_x\phi_i +\text{constant}
\end{align*}
Thus, up to total derivatives and constants which can be dropped, the above transformation eliminates the random potential term. Let us always assume that such a transformation has been carried out properly, thus Eq. \eqref{app_eq:2/3_Lagrangian_clean} describes the ``clean" edge in the possible presence of chemical potential disorders. The general disordered edge in the presence of a disorder due to a vertex operator then takes the form of Eq. \eqref{eq:full_L_general} in the main text. Moreover, given the basis rotation in Eq. \eqref{eq:transformation to remove chemical disorder} (which contains randomness from the random potential $\mu(x)$), even though the tunneling/pairing coefficient $\xi$ may be uniform in the original basis, $\xi\rightarrow\xi(x)$ becomes random in the rotated basis. This provides further motivation for us to focus on the random disorder effect of $\mathcal{O}_{q,p}$ as analyzed in the main text. \\

\subsection{Decoupling of counter-propagating modes}
Here we show that at the fixed point $g=g_{q_N,p_N}\equiv g_N$ (see Eq. \eqref{app_eq: g*}), the edge theory composes of two \textit{decoupled} counter-propagating boson modes. As with the main text, we focus on cases with $p_N^2-3q_N^2=1$, see Eq. \eqref{eq:p_q_relation}, which correspond to having a relevant disorder operator $\mathcal{O}_{q_N,p_N}$ acting on the edge to induce the SC$_N$ phase. Consider the change of basis in Eq. \eqref{eq: change_basis_charge_neutral}, i.e., $\begin{pmatrix}
        \phi_u \\ \phi_d
    \end{pmatrix}= \begin{pmatrix}
      p_N & q_N\sqrt{3} \\
       q_N\sqrt{3} & p_N 
    \end{pmatrix}\begin{pmatrix}
        \phi_\sigma \\ \phi_\rho
    \end{pmatrix}$,  
and use the definitions of $\phi_\rho$ and $\phi_\sigma$ in Eq. \eqref{eq: def charge and neutral basis}, it is straightforward to rewrite the Lagrangian in Eq. \eqref{app_eq:2/3_Lagrangian_clean} as 
\begin{equation}
    \mathcal{L}_0 = \frac{1}{4\pi}\partial_x\phi_u(\partial_t-v_u \partial_x)\phi_u  - \frac{1}{4\pi}\partial_x\phi_d(\partial_t+v_d\partial_x)\phi_d-\frac{v_{ud}}{2\pi}\partial_x\phi_u\partial_x\phi_d. 
\end{equation}
It is clear in this form that $\phi_u$ and $\phi_d$ are the respective upstream and downstream modes, satisfying $[\partial_x\phi_{u}(x),\phi_{u}(x')]=-2i\pi\delta(x-x')$, $[\partial_x\phi_{d}(x),\phi_{d}(x')]=2i\pi\delta(x-x')$ and $[\phi_u,\phi_d]=0$. The transformation on the velocity matrix gives:
\begin{equation}
\begin{split}
    &\bpm v_u & v_{ud} \\
    v_{ud} & v_d
    \epm = \frac{1}{\sqrt{2}}\left(
\begin{array}{cc}
 -p_N-3 q_N & \sqrt{3}(p_N+q_N) \\
 p_N+q_N & -\frac{1}{\sqrt{3}}(p_N+3 q_N) \\
\end{array}
\right)^T \bpm v_{11} &v_{12} \\ v_{12} & v_{22} \epm \frac{1}{\sqrt{2}}\left(
\begin{array}{cc}
 -p_N-3 q_N & \sqrt{3}(p_N+q_N) \\
 p_N+q_N & -\frac{1}{\sqrt{3}}(p_N+3 q_N) \\
\end{array}
\right) \\
&\implies v_{ud} =  -\frac{1}{2 \sqrt{3}}\left[ (3 v_{11}+v_{22})(p_N+q_N)(p_N+3q_N) -4v_{12}(p_N^2+3q_N^2+3p_Nq_N) \right]
\end{split}
\end{equation}
Upon using Eq. \eqref{eq: def_g} to replace $v_{12}$ by $g$, and using the expression for the fixed-point interaction parameter $g_{q,p}$ in Eq. \eqref{app_eq: g*}, we obtain
\begin{equation}
    v_{ud} = \frac{3v_{11}+v_{22}}{\sqrt{3}}\frac{(p_N+q_N)(p_N+3q_N)}{(g_N-1) (g+1)}(g-g_N), 
\end{equation}
for the SC$_N$ phase. Near the fixed point, $v_{ud}$ vanishes linearly as $v_{ud}\propto (g-g_N)$. As we have mentioned in the main text, it is not difficult to see (without any tedious algebraic manipulations) that at $g=g_N$ we must have decoupled up- and down-stream modes. This is because the operator $\mathcal{O}_{q_N,p_N} = e^{i\sqrt{2}\phi_u}$ is then completely chiral with its scaling dimension minimized to be equal to its conformal spin, which is 1. The derivation presented here serves to show in detail how $v_{ud}$ vanishes as the RG fixed-point is approached.\\

\section{Scattering analysis for quantized linear transport}\label{sec: scattering}
\setcounter{equation}{0}
\setcounter{figure}{0} 
Here we provide more details for the quantized linear transport calculation regarding the KFP-SC$_N$-KFP junction (the lower edge shown in main text \cref{fig:KFP-SC-KFP}). We first derive the boundary conditions at the KFP-SC$_N$-KFP junction interface, with an elaborate justification for performing a disorder-average for the boundary conditions leading to Eq. \eqref{eq:simple-bc}. We then provide a review of the Landauer-B\"{u}ttiker formula applied in the presence of superconducting leads. We derive Eq. \eqref{eq: Rd} and also explain how the four-terminal Hall resistance $R_H$ is insensitive to the distinction of different SC$_N$ phases, which in turn highlights the importance of the downstream resistance $R_d$.

\subsection{Boundary conditions}
To address the scattering problem, we need to specify the boundary condition at the junction interface where the SC$_N$ region meets the KFP region. Because the up/down-stream modes of different regions correspond to different rotations of the original basis $(\phi_1, \phi_2)$, the way to deduce the boundary condition is to go back to the original basis and analyze the corresponding equation of motion. The Lagrangian of interest is
\begin{equation}
    \mathcal{L} = -\frac{1}{4\pi}\sum_{ij} [K_{ij}\partial_t\phi_i \partial_x \phi_j + v_{ij}(x)\partial_x\phi_i \partial_x\phi_j]+ [\xi_{0}(x)\mathcal{O}_{0,1}+ \xi_N(x) \mathcal{O}_{q_N,p_N}+h.c.],
\end{equation}
where $\mathcal{O}_{q,p} \equiv e^{i(3q+p)\phi_1+i(3q+3p)\phi_2}$. We consider a spatially dependent velocity matrix $v_{ij}(x)$ such that it is discontinuous at the junction interface at $x=0$ and $x=L$, such that the corresponding interaction parameter $g(x)$, see Eq. \eqref{eq: def_g} and Eq. \eqref{app_eq: Lambda and g_def2}, takes different fixed-point values (c.f. Eq. \eqref{app_eq: gfixed}): $g(x)=g_0$ in the KFP regions $x<0$ and $x>L$, and $g(x)=g_N$ in the SC$_N$ region $0<x<L$. Correspondingly, the renormalization group flow (c.f. Sec. \ref{sec: RG}) is such that the disorder coupling $\xi_0(x)$ flows to zero for $0<x<L$ (where $\xi_N(x)$ flows to strong coupling), while $\xi_N(x)$ flows to zero for $x<0$ or $x>L$ (where $\xi_0(x)$ flows to strong coupling).

By the Euler-Lagrange equation, we obtain
\begin{equation} \label{app_eq: EOM}
    \sum_jK_{ij} \partial_t \partial_x\phi_j +\partial_x[v_{ij}(x)\partial_x\phi_j] = F_i(x) \equiv 
    \begin{cases}
        -2\pi i [\xi_0(x)\mathcal{O}_{0,1}+(3q_N+p_N)\xi_N(x)\mathcal{O}_{q_N, p_N}] + h.c.\;\;\;&(i=1)\\
        -6\pi i[\xi_0(x)\mathcal{O}_{0,1}+(q_N+p_N)\xi_N(x)\mathcal{O}_{q_N,p_N}]+h.c.\;\;\;&(i=2)
    \end{cases}\;.
\end{equation}
While $F_i(x)$ is not expected to be smooth or continuous, it is always finite. Integrating the equation of motion over $x$ leads to 
\begin{equation}
    \partial_x\phi_k = \sum_i v^{-1}_{ki}(x)\Big[\sum_j -K_{ij} \partial_t \phi_j + \int^x F_i(x') dx' + \text{const.}\Big]\ .
\end{equation}
Since we expect nothing divergent on the right, integrating both sides of the above expression over an infinitesimal region around the junction interface leads to the continuity of $\phi_{1,2}$ at the interface. In turn, it is also clear from directly integrating Eq. \eqref{app_eq: EOM} over an infinitesimal region about the interface that $\sum_j v_{ij}(x) \partial_x \phi_j$ should be continuous, because the continuity of $\phi_{1,2}$ implies the continuity of $\partial_t\phi_{1,2}$. Altogether, the boundary conditions are
\begin{subequations}\label{eq: scattering BC}
    \begin{align}
        \phi_i(x_*^-) &= \phi_i(x_*^+) \label{eq: scattering BC (a)} \\
        \sum_j v_{ij}(x_*^-) \partial_x \phi_j(x_*^-) &= \sum_j v_{ij}(x_*^+) \partial_x \phi_j(x_*^+) \label{eq: scattering BC (b)}
    \end{align}
\end{subequations}
where $x_*=0$ or $L$ denotes the position of the interface, and $v_{ij}$ is the velocity matrix evaluated at the RG fixed point of the corresponding region. 

In the main text, we have used an equivalent form of Eq. \eqref{eq: scattering BC (b)}, which is expressed in terms of the fixed-point basis. Without loss of generality, we consider the case where the edge is in the KFP phase at $x<x_*$, and in the SC$_N$ phase at $x>x_*$ (the opposite case follows a similar argument as below). To that end, let us define $(\phi_u, \phi_d)^T = R_{ud} (\phi_1, \phi_2)^T$ and $(\phi_\sigma, \phi_\rho)^T = R_{\sigma\rho} (\phi_1, \phi_2)^T$, such that the matrix $\mathcal{R}_{N}$ introduced in the main text in Eq. \eqref{eq: change_basis_charge_neutral} satisfies $\mathcal{R}_{N} = R_{ud}R^{-1}_{\sigma\rho}$. Notice that under this change of basis, the velocity matrix transformed as follows:
\begin{equation}
    \bpm v_{11}(x^-_*) & v_{12}(x^-_*) \\
    v_{12}(x^-_*) & v_{22}(x^-_*)\epm = R^T_{\sigma\rho} \bpm v_\sigma & 0 \\ 0 & v_\rho \epm R_{\sigma\rho}
    \quad\quad \text{and}\quad\quad  \bpm v_{11}(x^+_*) & v_{12}(x^+_*) \\
    v_{12}(x^+_*) & v_{22}(x^+_*)\epm = R^T_{ud} \bpm v_u & 0 \\ 0 & v_d\epm R_{ud}
\end{equation}
Then, Eq. \eqref{eq: scattering BC (b)} can be recast as
\begin{equation}
\begin{split}
     R^T_{\sigma\rho} \bpm v_\sigma \partial_x \phi_\sigma \\ v_\rho \partial_x \phi_\rho \epm_{x=x^-_*} = R^T_{ud} \bpm v_u \partial_x \phi_u \\ v_d \partial_x \phi_d \epm_{x=x^+_*} &\implies \bpm v_\sigma \partial_x \phi_\sigma \\ v_\rho \partial_x \phi_\rho \epm_{x=x^-_*} = \mathcal{R}_{N} \bpm v_u \partial_x \phi_u \\ v_d \partial_x \phi_d \epm_{x=x^+_*} \\
    & \implies  \mathcal{R}_{N} \bpm - v_\sigma \partial_x \phi_\sigma \\ v_\rho \partial_x \phi_\rho \epm_{x=x^-_*} = \bpm -v_u \partial_x \phi_u \\ v_d \partial_x \phi_d \epm_{x=x^+_*}.
\end{split}
\end{equation}
The last expression is what we used in the main text to obtain Eq. \eqref{eq:simple-bc}. It is also expressed in a more suggestive form when compared with Eq. \eqref{eq: scattering BC (a)}, $\partial_t\phi_i(x^-_*)=\partial_t\phi_i(x^+_*) $, which implies
\begin{equation}
    \mathcal{R}_N\bpm \partial_t \phi_\sigma \\ \partial_t \phi_\rho \epm_{x=x^-_*} =  \bpm \partial_t\phi_u \\ \partial_t\phi_d \epm_{x=x^+_*},
\end{equation}
and can be seen as consistent with the equation of motion for the free boson fields $\phi_\rho$ and $\phi_d$. 

\subsection{Disorder averaging}\label{app_sec: disorder averaging justification}

As explained in the main text, the boundary conditions obtained just above are further simplified to \cref{eq:simple-bc} upon disorder averaging, which sets the upstream particle current at the upstream end-point $\partial_x\phi_u\vert_{x=0^+} =0$. Here we provide a more elaborate justification for this treatment. Recall \cref{eq:rand-rot},
\begin{equation}
\partial_x\phi_u =n_z(x) \partial_x\widetilde{\phi}_u+\sqrt{2}\pi\Big[n_+(x) e^{-i\sqrt{2}\widetilde{\phi}_u}+h.c.\Big], 
\end{equation}
which relates $\phi_u$ to the free chiral boson $\tilde{\phi}_u$ moving upstream from $x=L$ to $x=0$. Here $U(x)\sigma_zU^\dagger(x) = n_z(x)\sigma_z+n_+(x)\sigma^+(x)+n_-(x)\sigma^-(x)$, and
\begin{equation}
    U(x) = \mathcal{P}_x \exp\{\frac{i}{v_u}\int^{x}_{x_0} \diff{x'}\; [\xi(x')\sigma^- + \xi^*(x')\sigma^+]\}
\end{equation}
is the random $SU(2)$ rotation introduced in the main text around Eq. \eqref{eq:SC_fixed point_L} to obtain the SC$_N$ fixed point. Setting the reference point for the rotation as $x_0=L$, we have $\partial_x\phi_u\vert_{x=L^-}=\partial_x\tilde{\phi}_u\vert_{x=L^-}$, while at the upstream end-point $\partial_x \phi_u\vert_{0^+}$ is the projection of the vector-current of the free mode $\tilde{\phi}_u$ onto a randomly rotated vector given by $n_{\mu=z,\pm}(0)$. For long enough $L$ (which we quantify below), the disorder-average of $n_{\mu=z,\pm}(0)$ vanishes, and so does the disorder-average of $\partial_x \phi_u\vert_{0^+}$.\\

Let us further justify the procedure for carrying out an average over the random disorder when deriving the boundary conditions. There are two reasons: (1) if the SC-proximity disorder is dynamically and randomly fluctuating at a time scale much shorter than the time of measurement, then the transport phenomenon is naturally self-averaging. (2) Even if the disorder is quenched, there is an intrinsic averaging effect due to the renormalization group flow. When we inject a current on the edge, the current consists of electrons in an energy window $[0, eV]$ and hence the edge is at a finite non-zero energy scale at which the RG-flow should be cut off. Consequently, $v_{ud}$ is not exactly zero and $v_u$ should have fluctuations at the scale of $v_{ud}$. Thus, $ U(x) = \mathcal{P}_x \exp\{\frac{i}{v_u}\int^{x}_{x_0} \diff{x'}\; [\xi(x')\sigma^- + \xi^*(x')\sigma^+]\}$, as well as $n_{z,\pm}(x)$, are energy-dependent and thus should be taken average over different electrons in the energy window $[0,eV]$, in which the corresponding upstream mode $\tilde{\phi}_u$ has different velocities $v_u$ and hence different configurations of $n_{z,\pm}$ are being sampled. For the sampling to be random, we require
\begin{equation}
    \frac{\delta v_u}{v_u}L \sim \frac{v_{ud}}{v_u}L \gg \frac{v_u^2}{W} \quad \implies L \gg V^{-1/2}
\end{equation}
where $v_u^2/W$ is the length scale beyond which $U(x)$ is uncorrelated, i.e.,  $\overline{U(x)U(x')}=0$ for $\abs{x-x'}\gg v_u^2/W$, where $W$ is the disorder strength such that $\overline{\xi(x)\xi^*(x')}=W\delta(x-x')$. From the RG-flow of $v_{ud}$ (see the main text discussion around \cref{eq:SC_fixed point_L}, as well as \cref{sec: RG} of this SM), when the edge is at an energy scale $eV$, we expect $v_{ud}\sim V^{1/2}$, and hence the disorder-averaging is valid for long junction length $L \gg V^{-1/2}$ (here we omit some proportionality factors related to the disorder strength as well as edge velocities). In this limit, the disorder-averaged boundary conditions in \cref{eq:simple-bc} are justified, and give a quantized charge transmission fraction $t_{eh}=1/p^2_N$.\\

As a final remark on the junction length $L$, let us remind the reader that there are higher-order corrections to $t_{eh}$ coming from the perturbatively irrelevant electron-vortex tunneling or Cooper-pair tunneling. This is discussed towards the end of the main text, and further elaborated in \cref{sec: nonlinear}. For such corrections $\delta t_{eh}$ to be small compared to the leading order quantized result, we require $ V^{-\alpha}\gg L$ (c.f. \cref{eq: Rd-non} of the main text, with $\alpha=1$ or 4 depending on whether electron-vortex tunneling dominates or Cooper-pair tunneling dominates). As $\alpha>1/2$, from the scaling analysis we expect there exists a reasonable regime for $L$ where the quantized transport can be observed.\\

\subsection{The Landauer-B\"{u}ttiker formula}\label{app_sec: LB formalism}
\subsubsection{Downstream resistance}
\begin{figure}
    \centering
    \resizebox{\columnwidth}{!}{\includegraphics[]{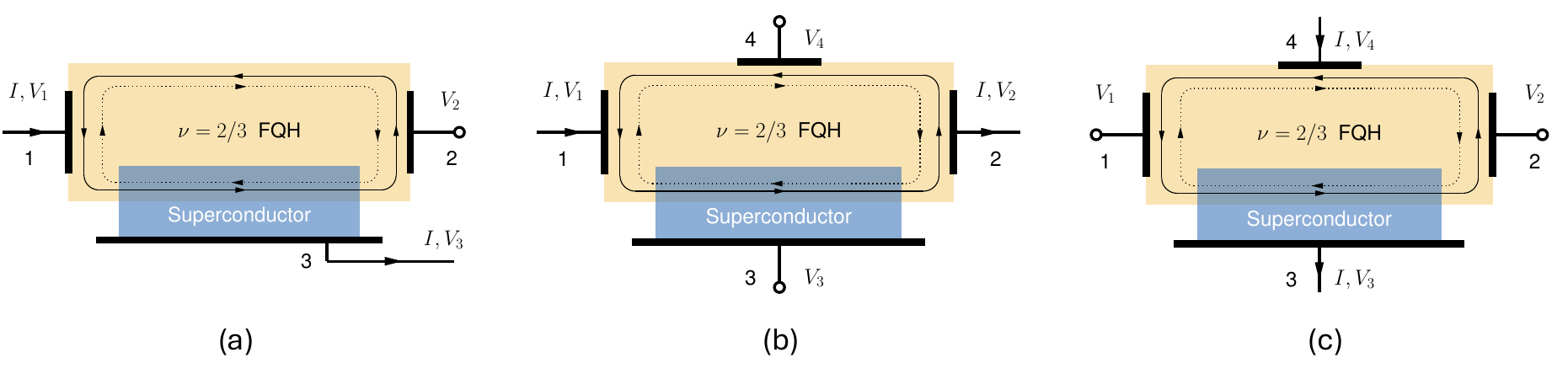}}
    \caption{Proposed experimental setup for studying quantized transport of the superconducting proximitized $\nu=2/3$ quantum Hall edges. (a) The three-terminal setup considered in the main text where the downstream resistance $R_d \equiv (V_2-V_3)/I$ is shown to be quantized when the proximitized edge is situated in the SC$_{N\neq 0}$ phase. (b,c) Four-terminal setups for probing the standard Hall resistance $R_H$ (with $R_H\equiv (V_3-V_4)/I$ for (b) and $R_H \equiv (V_1-V_2)/I$ for (c)). Irrespective of the nature of the proximitized edge, $R_H = 3h/(2e^2)$. The quantization of $R_d$ is thus a new and  distinguished type of quantized transport uniquely arising from the interplay between superconductivity and fractional quantum Hall edge physics. 
    }
    \label{supp_fig:LBsetup}
\end{figure}

The Landauer-B\"{u}ttiker formula relates the electrical transport of a mesoscopic quantum conductor to its scattering properties in a multi-terminal setup. In our setup shown in \cref{fig:setup} of the main text, as well as Fig. \ref{supp_fig:LBsetup}(a), there are two normal leads (labeled as 1 and 2) where individual electrons can enter or leave, and one superconducting lead (labeled as 3) where only Cooper pairs can enter or leave and hence making Andreev reflection processes essential. The generalized Landauer-B\"{u}ttiker formula, with the effect of Andreev processes in the presence of SC taken into account \cite{BTK1982, Anantram1996_AndreevScattering, Aharony2008, hu2024resistance}, reads as follows:
\begin{equation}
\left\{\begin{aligned}
    I_1 &= \frac{\nu e^2}{h}(V_1-V_2),\\
    I_2 &= \frac{\nu e^2}{h}[(V_2-V_3)-t_{eh}(V_1-V_3)],\\
    I_3 &= -I_1-I_2,
\end{aligned}\right.
\end{equation}
where $I_i\;( i=1,2,3)$ denotes the current flowing out of the lead $i$ into the sample. Here $\nu=2/3$ is the fractional filling of the FQH state, which is associated to the two-terminal quantized conductance $G=I_1/(V_1-V_2)=\frac{2e^2}{3h}$ of the chiral charge mode in the KFP phase (which is experimentally the case without SC proximity). The KFP phase is assumed to describe the entire FQH edge except for the region proximitized by the SC lead, which could lie in the SC$_N$ phase. The expression for $I_1$ thus takes the usual form for an ordinary Landauer-B\"{u}ttiker formula, with a chiral charge mode flowing from lead 2 to lead 1 (c.f. Fig. \ref{fig:setup}). As for $I_2$, there are two independent contributions. The first term is again the usual one due to the chiral charge mode flowing from lead 3 to lead 2. The second term is proportional to $t_{eh}$ as defined in \cref{SM-eq:teh} below (introduced in the main text, it is the transmission fraction of the charge going downstream through the SC-proximitized region), which has the interpretation of a current originating from lead 1, transmitting through the proximitized region, and flowing into lead 2. Finally, the expression for $I_3$ follows from the net current conservation. \\

In defining the downstream resistance in the setup depicted in the main text Fig. \ref{fig:setup} (see also Fig. \ref{supp_fig:LBsetup}(a)), lead 2 is a floating voltage probe such that $I_2=0$. Then, we have $I_1 = -I_3 = I$ and
\begin{equation}
    V_2-V_3 = t_{eh}(V_1-V_3)\Longrightarrow V_1-V_2 = (1-t_{eh})(V_1-V_3)
\end{equation}
which leads to the downstream resistance
\begin{equation}\label{app_eq: R_d from LB}
    R_d = \frac{V_{2}-V_3}{I} = \frac{h}{\nu e^2}\frac{t_{eh}}{1-t_{eh}}.
\end{equation}
This is the expression used in Eq. \eqref{eq: Rd} of the main text.\\

In anticipation of the subsequent discussion of nonlinear tunneling current, let us remark that the charge transmission fraction across the SC lead, $t_{eh}$, can be expressed as
\begin{equation}\label{SM-eq:teh}
    t_{eh} = t_n - t_a,\quad t_n+t_a =1
\end{equation}
where $t_n$ and $t_a$ are the normal (i.e., electron-to-electron) transmission probability and Andreev (i.e., electron-to-hole) transmission probability, respectively. 
The tunneling conductance to be discussed next, $\diff{I_T}/\diff{V_{13}}$, is proportional to $t_a = (1-t_{eh})/2$, which allows us to derive the nonlinear voltage dependence of the downstream resistance as in Eq. \eqref{eq: Rd-non}.\\

\subsubsection{Hall resistance}
For completeness, let us also calculate the Hall resistance in the presence of superconducting proximity using the four-terminal setup depicted in Fig. \ref{supp_fig:LBsetup}(b,c). The Landauer-B\"{u}ttiker formula reads as follows:
\begin{equation}
\left\{\begin{aligned}
    I_1 &= \frac{\nu e^2}{h}(V_1-V_4),\\
    I_2 &= \frac{\nu e^2}{h}[(V_2-V_3)-t_{eh}(V_1-V_3)],\\
    I_3 &= -I_1-I_2-I_4,\\
    I_4 &=\frac{\nu e^2}{h}(V_4-V_2).
\end{aligned}\right.
\end{equation}
The standard Hall resistance can be defined in two manners, one as in Fig. \ref{supp_fig:LBsetup}(b) with lead 3 and lead 4 being the floating voltage probes, and another as in Fig. \ref{supp_fig:LBsetup}(c) with lead 1 and lead 2 as the floating voltage probes. In (b), $I_3=I_4=0 \implies I_1=-I_2 \implies V_1-V_2 = -V_2+V_3+t_{eh}(V_1-V_3) \implies (1-t_{eh})(V_1-V_3)=0 \implies V_1=V_3$\footnote{For the case $t_{eh}=1$, that is when the proximitized edge is in the KFP phase just as the rest of the edge, it is also obvious that $V_1=V_3$.}, thus $R_H \equiv (V_3-V_4)/I_1 = 3h/(2e^2)$. Similarly, in (c) we have $I_1=0 \implies V_4=V_1$ and hence again $R_H \equiv (V_1-V_2)/I_4 = 3h/(2e^2)$. Therefore, $R_H$ is insensitive to $t_{eh}$ or the nature of the proximitized edge. This further supports the other transport quantity discussed above, i.e., the downstream resistance $R_d$, is associated to a new type of quantized transport beyond the quantum Hall effect that uniquely reflects the interplay between superconductivity and fractional quantum Hall edge physics. 

\section{Nonlinear transport calculation}\label{sec: nonlinear}
\setcounter{equation}{0}
\setcounter{figure}{0} 
In this section, we provide detailed explanations for deriving the nonlinear correction to the downstream resistance $R_d$ in Eq. \eqref{eq: Rd-non}. We first calculate in detail the effect of Cooper-pair tunneling between the FQH edge and the SC lead using perturbation theory, and then generalize the discussion to the tunneling of (multi-)electrons between the FQH edge and the near-edge vortices residing within the SC lead. Finite temperature effects are also discussed. The schematic setup for the following calculations is depicted in Fig. \ref{fig:KFP-SC-KFP}. 

\subsection{Cooper-pair tunneling}
The starting point of the nonlinear transport calculation is the perturbation Hamiltonian corresponding to the tunneling of Cooper-pair (or more generally, even-charged cluster) between the FQH edge and the superconductor (SC). In the Heisenberg picture, we have
\begin{equation}\label{app_eq: Cooper-pair H_T}
    H_T(t)=\int_0^L dx [e^{i\frac{2e}{\hbar}q_\text{eff}V_{13}t}\xi_{p_N,3q_N}(x)\mathcal{O}_{p_N,3q_N}(x,t) +h.c.].
\end{equation}
Here we have stipulated that the disordered FQH edge (along the interval $x\in[0,L]$, see also main text Fig. \ref{fig:KFP-SC-KFP}) has been driven to the SC$_N$ fixed point, so the nonlinear current is dominated by the Cooper pair tunneling with the least irrelevant scaling dimension, which we find is given by the operator $\mathcal{O}_{p_N, 3q_N}$ annihilating charge $-2p_Ne$, thus tunneling $p_N$ Cooper pairs (see Sec. \ref{app_sec: Cooper-pair scaling dimension}). We again assume a random, spatially uncorrelated, tunneling amplitude $\xi_{p_N, 3q_N}(x)$ with ensemble averages $\overline{\xi_{p_N, 3q_N}(x)}=0$ and $\overline{\xi_{p_N,3q_N}(x)\xi_{p_N,3q_N}^*(x')}=W_T\delta(x-x')$ , and $W_T$ denotes the disordered tunneling strength. The dynamical phase $e^{i\frac{2e}{\hbar}q_\text{eff}V_{13}t}$ generally accounts for the energy difference between the edge excitation and the Cooper-pair condensate in the SC as induced by the biased voltage $V_{13}$ between lead 1 and the SC lead 3 (hence proportional to $V_{13}$). The precise meaning of $q_\text{eff}$ will be clarified in each scenario to be discussed below. To leading order, $q_\text{eff}$ is spatially constant. Later in Sec. \ref{app_sec: decaying q_eff} we refine the analysis for $q_\text{eff}(x)$ decaying along the proximitized edge which takes into account the current and energy dissipation along the edge.\\

The tunneling current operator can be obtained via the Peierls' substitution. Consider separating the FQH edge and the SC lead by a distance $w$ (in the direction perpendicular to the edge, say, the $z$ direction), then in the presence of a vector potential a Peierls' phase is inserted into $H_T$, with $\mathcal{\xi}_{p_N,3q_N} \rightarrow e^{\frac{i}{\hbar}2p_NeA_zw}\mathcal{\xi}_{p_N,3q_N} $. The operator for the tunneling current (flowing into the SC) is then
\begin{equation}\label{app_eq: Cooper-pair I_T operator expression}
    \hat{I}_T(t)  = \left.\frac{\partial H_T}{\partial (A_zw)}\right|_{A_z=0} = i\frac{2p_Ne}{\hbar}\int_0^L dx [e^{i\frac{2e}{\hbar}q_\text{eff}V_{13}t}\xi_{p_N,3q_N}(x)\mathcal{O}_{p_N,3q_N}(x,t) -h.c.].
\end{equation}
Alternatively, the expression of $ \hat{I}_T(t)$ can be written as $\hat{I}_T=\frac{p_N}{q_{\text{eff}}V_{13}}\frac{\partial H_T}{\partial t}$ using $H_T$ in the Schr\"odinger picture (i.e. using a time-independent $\mathcal{O}_{p_N,3q_N}$).

We now use the Kubo formula to compute the leading-order tunneling current: $I_T = \frac{i}{\hbar}\int_{-\infty}^0 dt\overline{\braket{[H_T(t),\hat{I}_T(0)]}}$, where $\overline{\braket{\quad}}$ means evaluating the expectation value under the steady state in the absence of $H_T$ and then  averaged over the random ensemble. After straightforward simplifications, we obtain
\begin{equation}\label{app_eq: intermediate expression for I_T}
    I_T = \frac{2p_NeW_T}{\hbar^2}\int_{-\infty}^\infty dt \int dx\;e^{i\frac{2e}{\hbar}q_{\text{eff}}V_{13}t} \braket{[\mathcal{O}_{p_N,3q_N}(x,t),\mathcal{O}^\dagger_{p_N,3q_N}(x,0)]}. 
\end{equation}
Recall $\mathcal{O}_{p_N, 3q_N} = e^{i\sqrt{6}\phi_d}$ (c.f. Sec. \ref{app_sec: Cooper-pair scaling dimension}) is a primary field at the SC$_N$ fixed point with scaling dimension $\Delta_{p_N,3q_N}=3$, thus \cite{giamarchi2003quantum, fradkin2013field} 
\begin{equation}\label{app_eq: zero-T correlator}
    \langle \mathcal{O}_{p_N,3q_N}(x,t)\mathcal{O}^\dagger_{p_N,3q_N}(0,0)\rangle = \frac{1}{[2\pi i(v_d t -x-i0^+)]^{2\Delta_{p_N,3q_N}}},
\end{equation}
where $v_d$ is the velocity of the downstream free chiral boson at the fixed point. $\langle \mathcal{O}^\dagger_{p_N,3q_N}(0,0)\mathcal{O}_{p_N,3q_N}(x,t)\rangle$ can be obtained from the above by replacing $(x,t) \rightarrow (-x,-t)$, hence
\begin{align}
    \braket{[\mathcal{O}_{p_N,3q_N}(x,t),\mathcal{O}^\dagger_{p_N,3q_N}(x,0)]} = \Big(\frac{1}{2\pi iv_d}\Big)^{2\Delta_{p_N,3q_N}}\left[\frac{1}{(t - i0^+)^{2\Delta_{p_N,3q_N}}} - \frac{(-1)^{2\Delta_{p_N,3q_N}}}{(t +i0^+)^{2\Delta_{p_N,3q_N}}}\right] \label{app_eq: contour integral} \\
    \implies \int_{-\infty}^{+\infty} dt\;e^{i\frac{2e}{\hbar}q_{\text{eff}}V_{13}t} \braket{[\mathcal{O}_{p_N,3q_N}(x,t),\mathcal{O}^\dagger_{p_N,3q_N}(x,0)]} = \frac{1}{v_d(2\Delta_{p_N,3q_N}-1)!}\left(\frac{e q_\text{eff}}{\pi\hbar v_d}V_{13}\right)^{2\Delta_{p_N,3q_N}-1}.
\end{align}
Notice that $(-1)^{2\Delta_{p_N,3q_N}}=1$ as $\Delta_{p_N,3q_N} \in\mathbb{Z}$, so irrespective of how we close the contour (through the upper/lower hemisphere for $e q_{\text{eff}}V_{13}\gtrless0$) we always obtain the same result. As we will see, this changes for the vortex tunneling analysis discussed below. Substituting the above result back into Eq. \eqref{app_eq: intermediate expression for I_T}, we arrive at
\begin{equation}
    I_T = \gamma W_TLV_{13}^{2\Delta_{p_N,3q_N}-1},\quad \text{where }\gamma = \frac{2\pi \hbar^{-1}p_N (q_{\text{eff}})^{2\Delta_{p_N,3q_N}-1}}{(2\Delta_{p_N,3q_N}-1)!}\left(\frac{e}{\pi\hbar v_d}\right)^{2\Delta_{p_N,3q_N}}.
\end{equation}
The tunneling conductance is proportional to the Andreev transmission probability $t_a \propto dI_T/dV_{13}$, with $t_a = (1-t_{eh})/2$  based on the discussion in Sec. \ref{app_sec: LB formalism}. Hence, the irrelevant Cooper-pair tunneling contributes a correction to $t_{eh}$ as $ \delta t_{eh} \propto - W_TL\abs{V_{13}}^{\alpha}$, with $\alpha=2\Delta_{p_N,3q_N}-2$ ($\in2\mathbb{Z}$)\footnote{We choose to express our result in terms of  $\abs{V_{13}}^\alpha$ in anticipation of the vortex tunneling analysis, which gives an odd $\alpha$, see below.}. Below, we specify to the $N=0$ and $N\neq 0 $ cases (for the proximtized edge in the SC$_N$ phase), and particularly clarify the meaning of $q_{\text{eff}}$ in each case. 

\subsubsection{The $N=0$ case}
In this case, we simply have $q_{\text{eff}}=p_{0}=1$. This is because the proximitized edge remains in the KFP phase, so it only couples weakly to the SC lead via RG-irrelevant Cooper-pair tunneling. The entire FQH edge (more precisely the lower edge in main text Fig. \ref{fig:setup}, see also main text Fig. \ref{fig:KFP-SC-KFP}) can thus be treated as at voltage $V_{13}$ with respect to the SC lead. The energy difference between the condensate in the SC lead and the edge Cooper-pair is thus $2eV_{13}$. \\

Taking into account the Cooper-pair tunneling, the total $t_{eh} = 1/p_0^2 +\delta t_{eh}=1+\delta t_{eh}$. Following Eq. \eqref{app_eq: R_d from LB}, we obtain $R_d = \frac{h}{\nu e^2}\frac{1+\delta t_{eh}}{-\delta t_{eh}} \propto -1/\delta t_{eh} \propto W_T^{-1}L^{-1}\abs{V_{13}}^{-\alpha}$, with $\alpha=2\Delta_{p_N,3q_N}-2$, which is the second expression in main text Eq. \eqref{eq: Rd-non}.

\subsubsection{The $N\neq0$ case}
In this case, the FQH edge couples strongly to the SC lead via the RG-relevant term $\mathcal{O}_{q_N, p_N}$, and is driven to the SC$_N$ fixed point with a downstream free chiral boson $\phi_d$. When lead 1 is biased against lead 3 with a potential difference $V_{13}$, in a steady state, there is an in-coming electric current $\frac{\nu e^2}{h}V_{13} = \frac{e}{2\pi}\sqrt{\frac{2}{3}}\langle \partial_t \phi_\rho \rangle\vert_{0^-}  = -\frac{e}{2\pi}\sqrt{\frac{2}{3}} v_\rho \langle\partial_x \phi_\rho \rangle\vert_{0^-}$ from lead $1$. Using the scattering analysis discussed in the main text Eq. \eqref{eq:simple-bc}, we find the particle current of $\phi_d$ in the proximitized region $x\in[0,L]$ is
\begin{equation}\label{app_eq: I_d}
    I_d \equiv \langle\partial_t \phi_d \rangle\vert_{0^+} = -v_d \langle\partial_x\phi_d \rangle\vert_{0^+} = -\frac{1}{p_N}v_\rho \langle\partial_x\phi_\rho \rangle\vert_{0^-} = \frac{e}{p_N \hbar}\sqrt{\frac{2}{3}}V_{13}.
\end{equation}
The above provides a steady-state description for the proximitized FQH edge in the SC$_N$ phase in the presence of a voltage bias (before the higher order tunneling term $\mathcal{O}_{p_N, 3q_N}$ is considered), where the proximitized edge ($x\in[0,L]$) hosts a constant particle current $I_d$. This steady state can be transformed into an equivalent description with $\langle \partial_t\phi_d\rangle=0$ via a transformation: $\phi_d \rightarrow \phi_d+I_d t$, 
and hence $\mathcal{O}_{p_N,3q_N} = e^{i\sqrt{6}\phi_d} \rightarrow e^{i\sqrt{6}I_dt}\mathcal{O}_{p_N, 3q_N}$. Comparing with Eq. \eqref{app_eq: Cooper-pair H_T}, we obtain 
\begin{equation}\label{app_eq: q_eff vs I_d 1}
    \frac{2e}{\hbar}q_{\text{eff}}V_{13} = \sqrt{6}I_d \implies q_{\text{eff}} = \frac{1}{p_N}.
\end{equation}
After this transformation, one can perform the standard perturbation analysis (as we did earlier) with the unperturbed state satisfying $\langle \partial_t\phi_d\rangle=0$.

For $N\neq 0$, the leading contribution to $I_T$ is provided by the RG-relevant tunneling of $\mathcal{O}_{q_N,p_N}$, which is linear in $V_{13}$ and gives the quantized contribution to $t_{eh} = 1/p_N^2$. The least irrelevant tunneling of $\mathcal{O}_{p_N, 3q_N}$ provides a correction $ \delta t_{eh} = t_{eh}-1/p_N^{2} \propto - W_TL\abs{V_{13}}^{\alpha}$, with $\alpha=2\Delta_{p_N,3q_N}-2$, so again following Eq. \eqref{app_eq: R_d from LB}, we obtain $R_d - \frac{h}{2q_N^2e^2} \propto \delta t_{eh} \propto -W_TL\abs{V_{13}}^{\alpha}$, which is the first expression in main text Eq. \eqref{eq: Rd-non}.

\subsection{Vortex tunneling}\label{app_sec: nonlinear current: vortex tunneling}
Next, we generalize the above analysis to account for another source of higher-order nonlinear transport in our setup: electron tunneling between the FQH edge and near-edge vortices (created either by a magnetic field or by thermal activation). Multiple electrons ($Q$ electrons) can tunnel into a vortex simulaneously as a cluster (via a charge $Q$ vertex operator). In general, a cluster of even charge $Q$ can always be tunneled into the SC as Cooper-pairs, which would contribute a smaller tunneling exponent $\alpha$ than that if it tunnels into the vortex core (because hopping into the vortex core costs an extra scaling dimension from the vortex operator), and the effect has been already considered above. Therefore, we do not need to consider even charge $Q$ here, and in the below we focus on the tunneling of odd charge $Q$ into a vortex core. 

Here we first consider the case of tunneling into a single isolated vortex located at $x_0$, with a tunneling amplitude $\zeta_{Q,r}$. The corresponding tunneling Hamiltonian and tunneling current operator resemble Eqs. \eqref{app_eq: Cooper-pair H_T} and \eqref{app_eq: Cooper-pair I_T operator expression}, which read as follows:
\begin{subequations}
\begin{align}
    H_T(x_0,t) &= e^{i\frac{e}{\hbar}q_{\text{eff}}V_{13}t}\zeta_{Q,r}(x_0)F_Q^\dagger(x_0,t) \mathcal{K}_{Q,r}(x_0,t)+h.c. \label{app_eq: single vortex H_T}\\
    \hat{j}_T(x_0,t) &= i\frac{Qe}{\hbar}\left\{ e^{i\frac{e}{\hbar}q_{\text{eff}}V_{13}t}\zeta_{Q,r}(x_0)F_Q^\dagger(x_0,t) \mathcal{K}_{Q,r}(x_0,t)-h.c. \right\}
\end{align}
\end{subequations}
Here we are modeling a vortex as a quantum dot hosting a large number $N$ flavors of randomly coupled free fermions created by $f^\dagger_{j=1,2,....,N\gg1}$, and $F_Q^\dagger(x_0,t)=\prod_{j=1}^Q f_{j}^\dag$ above schematically represents a product of $Q>0$ such fermion operators (without loss of generality upon relabeling denoted as the first $Q$ fermion flavors), which has a total scaling dimension $Q/2$ (see Sec. \ref{app_sec: vortex model and RG}). The tunneling Hamiltonian with $Q=1$ is presented in the main text. The charge-$Qe$ edge cluster annihilation operator $\mathcal{K}_{Q,r}$ has been introduced before in Eq. \eqref{app_eq: odd-charge cluster operator}: $\mathcal{K}_{Q,r} = e^{iQ\sqrt{\frac{3}{2}}\phi_\rho +i\frac{2r-1}{\sqrt{2}}\phi_\sigma}$, with $Q\in 2\mathbb{Z}+1$ and $r\in\mathbb{Z}$ (which parametrizes the back-scattering), and its scaling dimension $\Delta_{\mathcal{K}_{Q,r}}$ has been studied in Sec. \ref{app_sec: scaling dimension: vortex tunneling}. Alternatively, written in terms of the fixed point basis of the SC$_N$ phase, we have
\begin{equation}
    \mathcal{K}_{Q,r} = e^{i\kappa_d \phi_d+i\kappa_u \phi_u}, \quad \text{with} \begin{cases}
        \kappa_d = \sqrt{\frac{3}{2}}[p_N Q-q_N(2r-1)] \\
        \kappa_u = \sqrt{\frac{1}{2}}[p_N(2r-1)-3q_NQ]
    \end{cases}
\end{equation}
Following again the analysis we did around Eq. \eqref{app_eq: I_d}, where we adopt an edge steady state for the SC$_N$ phase with $\langle\partial_t\phi_d\rangle=I_d =\frac{e}{p_N \hbar}\sqrt{\frac{2}{3}}V_{13}$, the precise meaning of $q_{\text{eff}}$ can then be read off as
\begin{equation}\label{app_eq: q_eff vs I_d 2}
    \frac{e}{\hbar} q_{\text{eff}}V_{13} = \kappa_dI_d \qquad \implies\qquad q_{\text{eff}} = \frac{\kappa_d}{p_N}\sqrt{\frac{2}{3}} = Q-\frac{q_N}{p_N}(2r-1).
\end{equation}
Now we can carry out the same Kubo formula calculation as performed in detail previously to obtain the leading-order tunneling current. Here we will be content with obtaining the tunneling exponent via power-counting. The single-vortex tunneling current is evaluated as $j_T(x_0) = \frac{i}{\hbar}\int_{-\infty}^0
dt\;\langle [H_T(x_0,t), \hat{j}_T(x_0,0)]\rangle$, thus
\begin{equation}\label{app_eq: j_T for a single vortex}
\begin{split}
    j_T(x_0)
&\propto \abs{\zeta_{Q,r}}^2 \int_{-\infty}^\infty dt\;e^{i\frac{e}{\hbar}q_{\text{eff}}V_{13}t} \langle[F_Q^\dagger(x_0,t)\mathcal{K}_{Q,r}(x_0,t), \mathcal{K}^\dagger_{Q,r}(x_0,0)F_Q(x_0,0)]\rangle \\
&\propto \abs{\zeta_{Q,r}}^2 \sgn(q_{\text{eff}} V_{13})(q_{\text{eff}}V_{13})^{\alpha+1} \propto  \abs{\zeta_{Q,r}}^2 \sgn(\kappa_d I_d)(\kappa_d I_d)^{\alpha+1}  \quad\text{with }\alpha = 2\Delta_{\mathcal{K}_{Q,r}}+Q-2\ ,
\end{split}
\end{equation}
where $\sgn(x)$ is the sign of $x$. The tunneling exponent $\alpha+1 = 2\Delta_{\mathcal{K}_{Q,r}}+Q-1$ can be simply read off by noticing that $\langle\mathcal{K}_{Q,r}(x_0,t) \mathcal{K}^\dagger_{Q,r}(x_0,0)\rangle \propto t^{-2\Delta_{\mathcal{K}_{Q,r}}}$, and $\langle F_Q^\dagger(x_0,t)F_Q(x_0,0)\rangle\propto \langle f_j^\dagger(x_0,t)f_j(x_0,0) 
\rangle^Q \propto t^{-Q}$ as consequence of modeling the vortex as a quantum dot (for more details, see Sec. \ref{app_sec: vortex model and RG}). Notice that in this calculation, in the counterpart  of \cref{app_eq: contour integral} we would have $(-1)^{2\Delta_{p_N, 3q_N}}$ there replaced by $(-1)^{2\Delta_{\mathcal{K}_{Q,r}}+Q} = -1$ (since $Q$ is odd and $\Delta_{\mathcal{K}_{Q,r}} \in\mathbb{Z}$ as explained in Sec. \ref{app_sec: scaling dimension: vortex tunneling}), which means the contour integral through the upper/lower hemisphere would differ by a sign, hence the factor $\sgn(q_{\text{eff}}V_{13})$ in the above equation.\\

In the presence of multiple vortices along the edge, as a leading order approximation, we can just add up individual vortex tunneling current by assuming the edge remains in the steady-state described by a constant $\partial_t\phi_d=I_d$ along the edge. Then, the net tunneling current is $I_T \propto W_T L \abs{V_{13}}^{\alpha}V_{13} \implies\delta t_{eh}\propto -dI_T/dV_{13}\propto-W_TL \abs{V_{13}}^\alpha$, with $W_T$ here proportional to both the vortex tunneling strength $\abs{\zeta_{Q,r}}^2$ and the density of the near-edge vortices. Similar to the analysis presented at the end of the last subsection, we arrive at the main text Eq. \eqref{eq: Rd-non} again but this time with $\alpha=2\Delta_{\mathcal{K}_{Q,r}}+Q-2$ ($\in2\mathbb{Z}+1$). Combined with the analysis presented in Sec. \ref{app_sec: scaling dimension: vortex tunneling}, we see that for $N=0,\pm 1$, the vortex tunneling effect dominates the nonlinear transport with $\alpha=1$ via a charge-$e$ tunneling ($Q=1$); for $N=\pm2$, the vortex tunneling effect dominates the nonlinear transport with $\alpha=3$ via a charge-$3e$ tunneling ($Q=3$); for $\abs{N}>2$, the vortex tunneling effect is never dominant (giving an exponent $\alpha\gg4$), and thus the Cooper-pair tunneling with $\alpha=4$ discussed earlier dominates the nonlinear transport.

\subsection{Refinement: spatial dependent $q_{\text{eff}}(x)$}\label{app_sec: decaying q_eff}
In all of the above analysis, we have assumed a constant $q_{\text{eff}}$, or equivalently, a constant particle current $\partial_t\phi_d=I_d$ (c.f. \cref{app_eq: q_eff vs I_d 1} and \cref{app_eq: q_eff vs I_d 2}) along the entire proximitized edge. This is only a leading order approximation, as Cooper-pair tunneling or vortex tunneling along the edge would lead to a decaying $q_{\text{eff}}(x) \propto I_d(x)$. To be specific, let us focus here on the case of vortex tunneling (which can be easily adopted to the case of Cooper-pair tunneling), with the decay of $I_d(x)$ related to the tunneling current (c.f. \cref{app_eq: j_T for a single vortex}) as 
\begin{equation}
    -j_T(x) \propto \frac{dI_d(x)}{dx} = -w_T \sgn(\kappa_d I_d) (\kappa_d I_d)^{\alpha+1} \implies \abs{I_d(x)}^{-\alpha} - \abs{I_d(0)}^{-\alpha} = \alpha \kappa_d w_T x,
\end{equation}
where $j_T(x)$ is the tunneling current flowing into the SC lead at $x$, its power-law scaling with $I_d(x)$ follows from the discussion in the previous subsections and $w_T$ is proportional to $W_T$ (proportional to the tunneling strength as well as the vortex density). The total tunneling current flowing into the SC is then
\begin{equation}
    I_T = I_d(0)-I_d(L) = I_d(0)[1-(1+\alpha w_TL\abs{I_d(0)}^\alpha)^{-\frac{1}{\alpha}}]\propto V_{13}[1-(1+\alpha W_TL\abs{V_{13}}^\alpha)^{-\frac{1}{\alpha}}],
\end{equation}
where we have used $I_d(0) = \frac{e}{p_N\hbar}\sqrt{\frac{2}{3}}V_{13}$ and rescaled $w_T$ to $W_T = w_T(\frac{e}{p_N\hbar}\sqrt{\frac{2}{3}})^\alpha$. We thus conclude from the above analysis that, as long as $W_TL\abs{V_{13}}^\alpha \ll 1$, we once again obtain $\delta t_{eh}\propto -dI_T/dV_{13}\propto -W_TL\abs{V_{13}}^\alpha$. Our analysis here provides the explicit condition for the main text \cref{eq: Rd-non} to hold. 

\subsection{Finite temperature effect}
In the above calculation, we have assumed the zero-temperature limit. The effect of finite temperatures becomes important when $k_B T \gg \abs{eV_{13}}$, which can be treated by replacing the zero-temperature correlator, say \cref{app_eq: zero-T correlator}, by its finite-temperature version: $\frac{1}{(v_d t-x-i0^+)} \rightarrow \frac{\pi k_B T/v_d}{\sinh\left[(\pi k_B T/v_d)(v_d t-x-i0^+)\right]}$ \cite{giamarchi2003quantum, fradkin2013field}, and repeating the above analysis correspondingly. Here instead, we adopt a more efficient argument following Ref. \cite{Fisher1994} that would lead to the same result regarding the power-law scaling with $T$. Notice that, for both the Cooper-pair tunneling (\cref{app_eq: Cooper-pair H_T}) and the vortex tunneling (\cref{app_eq: single vortex H_T}), the local tunneling term takes the form $\xi(x) \mathcal{O}(x)+h.c.$, where $\xi(x)$ is the tunneling amplitude and $\mathcal{O}(x)$ is the tunneling operator (i.e., $\mathcal{O}_{p_N,3q_N}$ for Cooper-pair tunneling and $(f^\dagger)^Q\mathcal{K}_{Q,r}$ for vortex tunneling). Consider a RG transformation integrating out a shell of frequencies between $e^{-\lambda}\Lambda$ and $\Lambda$ (an ultraviolet cutoff), the leading order RG flow equation is \footnote{Unlike in Sec. \ref{sec: RG} where we are interested in the (1+1)-d edge theory, here we do not rescale the spatial direction as the modeled tunneling is spatially uncorrelated. The total tunneling process along the entire edge is just a sum of tunnelings at each point.}: $\frac{\diff \xi}{\diff \lambda} = (1-\Delta)\xi \implies \xi(\lambda) \propto e^{-\lambda(\Delta-1)}$
where $\Delta$ is the scaling dimension of the tunneling operator $\mathcal{O}$. At finite temperatures ($k_B T \gg \abs{eV_{13}}$) the RG flows are cut off at the energy scale of $k_BT \sim e^{-\lambda}\Lambda$ \footnote{On the other hand, in the zero-temperature limit with $k_B T \ll \abs{eV_{13}}$, the RG flow cutoff is instead at $\abs{eV_{13}}$, and we recover the voltage-dependent tunneling probability as analyzed in detail in the previous subsections.}, thus the effective temperature-dependent tunneling amplitude is $\xi(T) \propto T^{\Delta-1}$. The tunneling strength, and the directly related Andreev transmission probability $t_a$ and the charge transmission fraction $\delta t_{eh}$, thus obey $\delta t_{eh} \propto -t_a\propto- \abs{\xi(T)}^2\propto T^{\alpha}$ ($\alpha=2\Delta-2$), which is what we have claimed in the main text around \cref{eq: Rd-non}. \\

\section{Vortex Model}\label{app_sec: vortex model and RG}
\setcounter{equation}{0}
\setcounter{figure}{0} 
In this final section, we provide details on how we model near-edge vortices as (0+1)-$d$ quantum dots with constant density of states, and argue that the coupling between the FQH edge and an isolated vortex is irrelevant in the renormalization group sense, which provides further justification to the perturbative calculation of the vortex tunneling current presented in the previous section. \\

\subsection{Vortex as a quantum dot}
The key feature we are invoking for a vortex (as a reservoir of normal electrons) is that it provides a nearly constant density of states for tunneling. Recall the electron density of states $\rho(E) \propto \text{Im}\;G^\tau(i\omega \rightarrow E+i0^+)$ with the Matsubara Green's function $G^\tau(i\omega) = \int_0^\infty d\tau G(\tau) e^{i\omega\tau}$, a constant density of states thus suggests $G(\tau) \propto 1/\tau$ (we use the imaginary-time formalism here for convenience, with $\tau=it$ and $t$ is the real time). A single vortex electron mode thus has a scaling dimension of $1/2$.  As such, a particular vortex model can be realized using a chiral free fermion, as adopted in Ref. \cite{Schiller2023}, and in that case an electron cluster of charge $Qe$ would have a scaling dimension of $Q^2/2$, which highly suppresses multi-electron tunneling to a near-edge vortex. Here we describe an alternative, arguably more realistic, vortex model which hosts charge-$Qe$ electron-cluster with a scaling dimension of $Q/2$.\\

We adopt the following action to describe a vortex:
\begin{equation}
    S_{\text{vortex}} = \int \diff{\tau}\Big[ \sum_{i=1}^N  f^\dagger_i \partial_\tau f_i - \sum_{i,j}f^\dagger_i M_{ij} f_j \Big],\;\;\text{with}\;\;\overline{M_{ij}} =0 \;\; \text{and}\;\; \overline{M_{ik}M_{lj}} = \delta_{ij} \delta_{lk}\frac{J^2}{N}.
\end{equation}
This is a statistical ensemble of $(0+1)$-dimensional quantum dots hosting a large number $N$ flavors of fermions ($f_{i=1,2,...,N\gg 1}$) whose bilinear mass terms are characterized by an $N\times N$ random hermitian matrix ($M_{ij}$) that obeys a Gaussian distribution (with the overline again representing the ensemble average).
The imaginary time-ordered fermionic Green's function, $G_{ij}(\tau) \equiv -\langle T_\tau f_i(\tau)f^\dagger_j(0)\rangle$, evaluates to the following upon ensemble averaging:
\begin{equation}
\begin{split}
    \overline{G_{ij}(\tau)} & = \int\frac{\diff{\omega}}{2\pi} \overline{(i\omega\mathds{1}-M)^{-1}_{ij}} e^{i\omega \tau} = \delta_{ij} \int\diff{E}\;\rho(E) \int\frac{\diff{\omega}}{2\pi} \frac{e^{i\omega \tau}}{i\omega -E}=\delta_{ij}\sgn(\tau) \int\diff{E}\;\rho(E) \theta(-E\tau)e^{E\tau}
\end{split}
\end{equation}
where $\theta(x)$ is the Heaviside step function, and $\rho(E)$ is the density of eigenvalues for random Hermitian matrices. It is well-known that, for large $N$, $\rho(E)$ obey the Wigner semicircle law \cite{zee2010quantum}: $\rho(E) = \frac{1}{2\pi J^2}\sqrt{4J^2-E^2}$ (note also that our $S_{\text{vortex}}$ is similar to the $q=2$ SYK model considered in Ref. \cite{PhysRevD.94.106002}, with their Majorana fermions there replaced by complex fermions here). In particular, the distribution near the eigenvalue zero (which amounts to the chemical potential of our vortex) is flat and given by the constant $\rho(0) = 1/(\pi J)$. Treating this constant density of states near $E=0$ as a generic characteristic of the modeled vortex bath, and considering the large $J\abs{\tau} \gg 1$ limit, we can approximate the ensemble-averaged Green's function as
\begin{equation}\label{app_eq: vortex Green's function}
    \overline{G_{ij}(\tau)} = \delta_{ij} \frac{\rho(0)}{\tau} = \delta_{ij} \frac{1}{\pi J \tau}.
\end{equation}
As such, we have effectively obtained $N$ decoupled flavors of fermionic modes in the vortex, each with a scaling dimension of $1/2$.

\subsection{Edge-vortex tunneling and RG analysis}

The tunnel coupling between the edge and the vortex is described by
\begin{equation}
    S_T  = \int \diff{\tau}\; \zeta_{Q,r}F_Q^\dagger \mathcal{K}_{Q,r} + h.c.
\end{equation}
with $\zeta_{Q,r}$ the tunneling amplitude and $F_Q^\dagger(t)=\prod_{j=1}^Q f^\dag_j(t)$ is a product of creation operators of $Q$ different fermionic modes of the vortex, which has a scaling dimension of $Q/2$ based on \cref{app_eq: vortex Green's function}. As introduced before in \cref{app_eq: odd-charge cluster operator}, $\mathcal{K}_{Q,r}$ annihilates an edge electron cluster of charge $Qe$ and has a scaling dimension of $\Delta_{\mathcal{K}_{Q,r}}$ (see \cref{app_sec: scaling dimension: vortex tunneling}). Carrying out an RG transformation for this term leads to the leading-order RG flow equation: $\frac{\diff \zeta_{Q,r}}{\diff \lambda} = (1-\Delta_{\mathcal{K}_{Q,r}}-\frac{Q}{2})\zeta_{Q,r}$, thus the charge-$Qe$ vortex tunneling can be RG-relevant only if $\Delta_{\mathcal{K}_{Q,r}} <1 - \frac{Q}{2} \leq \frac{1}{2}$. From the discussion in \cref{app_sec: scaling dimension: vortex tunneling}, we have seen that $\Delta_{\mathcal{K}_{Q,r}} \geq 1$, thus the tunnel coupling between the FQH edge and an isolated near-edge vortex is RG-irrelevant.

\end{document}